\documentclass[aps,prl,twocolumn,superscriptaddress,nofootinbib,floatfix]{revtex4-2}
\usepackage[T1]{fontenc}
\usepackage{amsmath,amssymb,bm,slashed}
\usepackage{graphicx,xcolor,booktabs,capt-of}
\usepackage{tikz}
\usetikzlibrary{arrows.meta,positioning}
\usepackage[hidelinks]{hyperref}
\usepackage[scr=rsfs]{mathalpha}

\usepackage[normalem]{ulem}     

\newcommand \widebar [1] {\overline{#1}}

\newcommand{\as}{a_s}
\newcommand{\dd}{ d}
\newcommand{\cusp}{\Gamma_{\mathrm{cusp}}}
\newcommand{\MS}{\overline{\mathrm{MS}}}

\definecolor{myblue}{rgb}{.8, .8, 1}
\definecolor{shadecolor}{rgb}{0.92,0.9,0.9}
 \definecolor{light-gray}{rgb}{0.827451,0.827451,0.827451}

\def \be  {\begin{equation}}
\def \ee  {\end{equation}}
\def \ba  {\begin{eqnarray}}
\def \ea  {\end{eqnarray}}
\def \baa {\begin{eqnarray*}}
\def \eaa {\end{eqnarray*}}
\def \lab #1 {\label{#1}}

\def\d{\hbox{{d}\kern-.20em\hbox{l}}}
\def \qqquad {\qquad\quad}
\def \qqqquad {\qquad\qquad}
\def \matrix #1 {\left(\begin{array}{cc} #1 \end{array}\right)}

\def \e  {\mathop{\rm e}\nolimits}

\def\II{\hbox{{1}\kern-.25em\hbox{l}}}

\def \H{\mathop{\rm H}\nolimits}

\def\inbar{\,\vrule height1.5ex width.4pt depth0pt}
\def\IC{\relax\hbox{$\inbar\kern-.3em{\rm C}$}}
\def\IZ{\relax{\hbox{\cmss Z\kern-.4em Z}}}
\def\IR{{\hbox{{\rm I}\kern-.2em\hbox{\rm R}}}}

\def\IP{{\hbox{{\rm I}\kern-.2em\hbox{\rm P}}}}
\def\II{\hbox{{1}\kern-.25em\hbox{l}}}

\begin{document}

\title{Three-Loop Evolution for Soft Functions of  Exclusive  $B$-Meson and Higgs Decays}

\date{\today}

\author{Yao Ji}
\email{yaoji@cuhk.edu.cn}
\affiliation{School of Science and Engineering, The Chinese University of Hong Kong, Shenzhen, Guangdong 518172, China}
\affiliation{Southern Center for Nuclear-Science Theory, Institute of Modern Physics, Chinese Academy of Sciences, Huizhou, Guangdong 516000, China}
\author{Bo-Xuan Shi}
\email{shibx@pku.edu.cn}
\affiliation{Center for High Energy Physics, Peking University, 5 Yiheyuan Road, Beijing 100871, China}
\author{Xing Wang}
\email{wangxing@cuhk.edu.cn}
\affiliation{School of Science and Engineering, The Chinese University of Hong Kong, Shenzhen, Guangdong 518172, China}
\affiliation{Southern Center for Nuclear-Science Theory, Institute of Modern Physics, Chinese Academy of Sciences, Huizhou, Guangdong 516000, China}
\author{Yu-Ming Wang}
\email{wangyuming@nankai.edu.cn}
\affiliation{School of Physics, Nankai University, Weijin Road 94, Tianjin 300071, P.R. China}

\begin{abstract}

We bootstrap the complete three-loop Lange--Neubert  kernel for the leading-twist $B$-meson light-cone distribution amplitude in heavy quark effective theory by taking advantage of the conformal symmetry technique. In particular, we provide for the first time  an all-order Wilson-line argument for the determination of the constant term in the LN kernel in terms of the light-quark, heavy-quark, and threshold soft anomalous dimensions. Remarkably,  this innovative construction requires no additional three-loop diagrammatic calculations. Taking into account further the three-loop QCD evolutions for the hard and hard-collinear coefficient functions, the newly derived three-loop Lange--Neubert equation enables us to manifestly demonstrate the renormalization-group invariance of the soft-collinear factorization formula for the radiative leptonic $B_{u}^- \to  \gamma \ell^{-} \bar \nu_{\ell}$  decay at leading power,   validating our bootstrapped three-loop kernel and more importantly, our full bootstrap approach.

\end{abstract}
\maketitle

\textit{Introduction.}---The light-cone distribution amplitudes (LCDA) of the bottom-meson  in heavy quark effective theory (HQET) \cite{Isgur:1989vq,Eichten:1989zv,Grinstein:1990mj,Georgi:1990um} play a central role in the field-theoretical description of exclusive $B$-meson decays into energetic particles.
Defined as hadronic matrix elements of the effective light-ray quark-gluon  operators \cite{Grozin:1996pq}, they encode the intricate strong interaction dynamics from the soft-scale fluctuation of the $B$-meson system.
The renormalization-scale dependence of these fundamental distribution functions can however be determined in perturbation theory. 
In particular, the leading-twist $B$-meson LCDA $\phi_B^{+}(\omega, \mu)$ turns out to be  indispensable for establishing QCD factorization theorems of  heavy-to-light $B$-meson decay form factors \cite{Beneke:2000wa,Beneke:2005gs,Hill:2004if,Becher:2004kk},  exclusive radiative and electroweak penguin $B$-meson decays \cite{Beneke:2001at,Beneke:2004dp,Becher:2005fg,Beneke:2020fot,Shen:2020hfq,Qin:2022rlk}, and two-body charmless $B$-meson decays at leading power in the heavy quark expansion \cite{Beneke:1999br,Beneke:2000ry,Beneke:2001ev,Beneke:2003zv,Lu:2022kos}.
Moreover, this HQET distribution amplitude is a key input to light-cone sum rules of a wide range of exclusive hadronic quantities \cite{DeFazio:2005dx,DeFazio:2007hw,Wang:2015vgv,Lu:2018cfc,Gao:2019lta,Wang:2017jow,Gao:2021sav,Cui:2022zwm,Cui:2023jiw} (see \cite{Khodjamirian:2023wol} for an overview) in soft-collinear effective theory (SCET).
Advancing our understanding towards both perturbative features and non-perturbative behaviours of  the leading-twist $B$-meson LCDA has therefore become the top priority, on the one hand,  to accomplish the model-independent and fully quantitative predictions of exclusive $B$-meson decay observables at an unprecedented level, and on the other hand, to decipher the ultimate mystery of numerous  persistent flavour anomalies observed at the LHCb and Belle II experiments (see, e.g., \cite{Capdevila:2023yhq} for further discussions on the current status).

An attractive non-perturbative determination of the leading-twist $B$-meson LCDA  with controllable theoretical uncertainties has  already  been proposed by measuring  the partially integrated branching fractions for the radiative leptonic $B$-meson decay ~\cite{Korchemsky:1999qb,Descotes-Genon:2002crx,Lunghi:2002ju,Bosch:2003fc,Beneke:2011nf,Braun:2012kp,Wang:2016qii,Wang:2018wfj,Beneke:2018wjp,Galda:2022dhp} and the exclusive four-body leptonic $B$-meson decay \cite{Beneke:2021rjf,Wang:2021yrr}.
The first-principles methodology of extracting  the whole profile of the leading-twist  $B$-meson LCDA  has  been  further  developed by carrying out the numerical simulation of an appropriate quasi-distribution amplitude in the framework of large momentum effective theory \cite{Wang:2019msf,Han:2024fkr,LatticeParton:2024zko,LPC:2026ffe,LPC:2026vyv}.
In sharp contrast with the renormalization-group (RG) evolution of the  collinear   pion LCDA in QCD, the normalization integral of the leading-twist  HQET $B$-meson distribution amplitude $\phi_B^{+}(\omega, \mu)$ becomes ultraviolet divergent at large values of $\omega$ \cite{Lange:2003ff,Braun:2003wx}.
Furthermore,  the asymptotic behaviour of this intriguing effective LCDA at large partonic momentum  can  develop  the negative radiation tail at an arbitrary renormalization scale \cite{Lee:2005gza,Braun:2003wx,Feldmann:2014ika,Beneke:2023nmj},  by virtue of the operator-product-expansion (OPE) analysis for the  regularized moments of $\phi_B^{+}(\omega, \mu)$.

A related problem occurs at much higher energy. The bottom-quark loop in $h\to\gamma\gamma$ is suppressed by the quark mass but contains large logarithms of $m_h^2/m_b^2$~\cite{Kotsky:1997rq}. Factorization at this subleading power separates hard, radiative jet, and soft-quark functions, with endpoint subtractions needed to define their convolutions~\cite{Liu:2019oav,Liu:2020wbn}. Their scale evolution resums the large logarithms~\cite{Liu:2020tzd}; the jet and soft functions have been calculated independently~\cite{Liu:2020ydl,Liu:2020eqe}. The soft-quark evolution is related to the LN kernel to all orders~\cite{Beneke:2024cpq} due to the ``double-copy'' relation at operator level. Thus a kernel originally developed for heavy-meson structure also controls logarithms in the Higgs-decay amplitude. The diphoton channel is used to measure Higgs production rates and constrain its couplings at the LHC~\cite{ATLAS:2022tnm,CMS:2021kom}.

The one-loop Lange-Neubert (LN) kernel was first derived in momentum space~\cite{Lange:2003ff} with its position-space counterpart followed soon after~\cite{Braun:2003wx}. Its conformal properties were later discovered~\cite{Braun:2014owa}, building on the previous study demonstrating that the eigenfunction of the one-loop LN kernel takes a simple form~\cite{Bell:2013tfa}. Subsequent studies employing the conformal techniques further revealed that the higher-twist $B$-LCDAs renormalization group equations (RGEs) also form completely integrable systems at the large-$N_c$ limit~\cite{Braun:2015pha,Braun:2017liq}.
The application of conformal symmetry in determining the LN kernel was then extended beyond one loop~\cite{Braun:2019wyx},  
by exploiting the Wilson-Fisher critical point in noninteger dimensions~\cite{Braun:2013tva,Braun:2014vba}, where conformal symmetry of QCD is restored~\cite{Braun:2018mxm}. This conformal approach, which was first pioneered in~\cite{Mueller:1993hg,Belitsky:1999gu} and later revived in~\cite{Braun:2013tva} from a different perspective, has determined the quantum-corrected conformal generators to two loops~\cite{Braun:2016qlg}, from which the three-loop anomalous dimension matrix of the nonforward twist-two nonsinglet operator was constructed to finite size~\cite{Braun:2017cih}, and the two-loop coefficient functions for deeply virtual Compton scattering was obtained~\cite{Braun:2020yib}. 

In this Letter, we bootstrap the complete three-loop LN kernel using the conformal approach, together with known anomalous dimensions as inputs. We first derive the two-loop corrections to the special conformal generator of the light-heavy operator from its light-light counterpart, hence determining the nonlocal part of the LN kernel. The remaining local constant is then fixed from the light-quark, heavy-quark, and threshold anomalous dimensions. Together, these ingredients establish the three-loop LN kernel without evaluating any Feynman diagrams. Exact $\epsilon$-pole cancellation in the independently calculated three-loop radiative jet function provides highly nontrivial validation for our bootstrapped LN kernel.

\providecommand*{\diphotoncontour}{%
  \tikz[x=3.3pt,y=3.3pt,baseline=0pt]{%
    \draw[line width=.45pt,line join=round]
      (0,0)--(0,1.12)--(.9,1.12)--(.9,.22)--(1.51,.22);%
  }%
}

\textit{Symmetry construction.}--- We derive the evolution kernel of the twist-two nonlocal light-heavy operator
\begin{equation}
 O(z)=\bar q(zn)\, \slashed{n}\gamma_5[zn,0]h_v(0),
 \label{eq:operator}
\end{equation}
from symmetry considerations. Here $q$ is the light-quark field, $h_v$ is of HQET~(see \cite{Neubert:1993mb,Manohar:2000dt} for excellent reviews) with timelike velocity $v$ satisfying $v^2=1$. $[zn,0]$ is a straight Wilson line in the lightlike direction $n^\mu$ so that $n^2=0$. We adopt the convention $n\cdot v=1$. 
Sandwiching $O(z)$ between the $B$-meson and vacuum states and dividing by the HQET decay constant $F_B$ defines the leading-twist $B$-LCDA $\Phi_+(z)$, the Fourier transform of $\phi_B^+(\omega)$. 
The LN kernel $\mathcal H$ governs the scale dependence of $O(z)$ through the RGE, \begin{align}
    \mu\frac{d}{d\mu}O(z)+[{\mathcal H}O](z)=0\, . 
\end{align}
The evolution kernel for $\Phi_+$ is therefore obtained by simply subtracting the anomalous dimension of $F_B$ already known to four loops~\cite {Chetyrkin:2003vi,Grozin:2023dlk}. The explicit definitions and relevant anomalous dimensions through three loops are provided in the Supplemental Material. 
The HQET field $h_v$ is equivalent to a semi-infinite timelike Wilson line stretching from $\infty v$ to $0$. By connecting to the finite lightlike link $[zn,0]$, it forms the cusp generating the explicit $\Gamma_{\rm cusp}\ln(i\widetilde\mu z)$ dependence~\cite{Braun:2003wx} in the position-space LN kernel with $\Gamma_{\rm cusp}$ being the universal cusp anomalous dimension and $\widetilde\mu\equiv\mu_{\widebar{\rm MS}}\e^{\gamma_E}$.

We derive the LN kernel by exploiting its properties under conformal transformations, see~\cite{Braun:2003rp} for an earlier overview on the conformal technique in QCD. This technique allows one to obtain the LN kernel directly at one loop, without any diagrammatic calculations, up to an additive (local) constant term~\cite{Braun:2014owa}. 
Beyond the leading order (LO), the application of the conformal technique in computing the LN kernel is obstructed by the nonvanishing QCD $\beta$-function in $d=4$ dimensions, as it explicitly breaks conformal symmetry in all aspects. 
However, this obstacle can be overcome by working at the Wilson-Fisher critical point of QCD in $d=4-2\epsilon_*$ with  $\epsilon_*$ fine-tuned so that the $\epsilon$-dependent $d$-dimensional $\beta$-function vanishes~\cite{Braun:2013tva,Braun:2014vba,Braun:2018mxm}, hence restoring the applicability of the conformal technique in obtaining the evolution kernel. 
The $\widebar{\rm MS}$-scheme LN kernel in four dimensions can then be reconstructed from its critical-point counterpart by trading $\epsilon_*$ with the four-dimensional $\beta$-function following the simple algebraic relation $\epsilon_*=-\beta_0a_s-\beta_1a_s^2+\cdots$. 
At the critical point, conformal symmetry dictates that 
\begin{equation}
 [\mathcal K,\mathcal H]=0,\quad
 [{\cal D},\mathcal H]=\cusp,
 \quad {\cal D}=z\partial_z+\tfrac32-\epsilon_* + {\cal H}.
 \label{eq:commutators}
\end{equation}
Hereafter, we suppress the $a_s=\alpha_s/(4\pi)$ dependence of these operators for simplicity. ${\cal K} = v^\mu {\cal K}_\mu$ with $ {\cal K}_\mu$ being the special conformal generator, and
\begin{align}
    {\cal K} = z[z\partial_z+2+\Delta]\, .\label{def:K}
\end{align}
It commutes with the LN kernel as shown in Eq.~\eqref{eq:commutators} because in the $v^\mu$ direction, the timelike Wilson line, i.e., the heavy quark field is invisible to the special conformal generator.
$\Delta$ in Eq.~\eqref{def:K} denotes the quantum correction to the canonical/classical ${\cal K}^{(0)} = z^2\partial_z+2z$.   
${\cal D}$ is the dilation operator, and the nonzero $[{\cal D},{\cal H}]$ in eq.~\eqref{eq:commutators} is a consequence of the semi-infinite timelike Wilson line explicitly breaking the scale invariance even at the critical point. 
On the other hand, the nonvanishing commutator at $d=4-2\epsilon_*$ given above is completely determined by the cusp anomalous dimension $\Gamma_{\rm cusp}$ thanks to its universality in reflecting the QCD double-pole structure~\cite{Korchemsky:1987wg}.
Together with the $SL(2,R)$ algebra $[{\cal D},{\cal K}] = {\cal K}$, we find the conformal representation of the LN kernel to all orders in perturbation theory~\cite{Braun:2019wyx},
\begin{equation}
 \mathcal H(\as)=\cusp(\as)
 \ln\!\left[i\mu e^{2\gamma_E}\mathcal K(\as)\right]
 +\Gamma_+(\as)\, .
 \label{eq:conformal}
\end{equation}
Equivalently, in position space the LN kernel takes the form, to all orders,
\begin{align}
 [\mathcal H f](z)&=\cusp\bigg\{
 \int_0^1\dd \alpha\,\frac{\bar \alpha}{\alpha}h(\alpha)
 [f(z)-f(\bar \alpha z)]\notag\\
 &\qquad + \ln(i\mu e^{\gamma_E}z)f(z)\bigg\}+\gamma_+(\as)f(z)\, ,
 \label{eq:kernel}
\end{align}
where $f(z)$ is an auxiliary test function and $\bar{\alpha} = 1-\alpha$.  Since $\cusp$ starts at $a_s$, the LN kernel is completely determined by ${\cal K}$ of lower $a_s$ orders up to an additive constant $\Gamma_+$. 
The three-loop LN kernel therefore requires $\mathcal K$ ($\Gamma_+$) to two (three) loops. Thus, our first task is to determine $\Delta^{(2)}$ in $\Delta=\as\Delta^{(1)}+\as^2\Delta^{(2)}+\cdots$.

The position-space light-to-heavy reduction procedure, which was introduced in Ref.~\cite{Braun:2018fiz} and applies to both the evolution kernel and the quantum corrections to the conformal generators, provides a convenient way of obtaining $\Delta^{(2)}$ from its counterpart $\Delta^{(2)}_{qq}$ calculated in Ref.~\cite{Braun:2016qlg} for the twist-two light-light operator 
\begin{align}\label{def:Oqq}
    O_{qq} = \bar q(z_1n)\slashed{n}[z_1n,z_2] q(z_2n)\, .
\end{align} 
Identifying $z_2$ as the location of the heavy field, the reduction procedure corresponds to neglecting all contributions in $\Delta_{qq}$ that shift the position of the second field as required by the heavy-quark limit. 
Here $\Delta_{qq}$ is the quantum correction to the special conformal generator associated with the light-light operator $O_{qq}$ in eq.~\eqref{def:Oqq}. 
This procedure reproduces the known one-loop result $\Delta^{(1)}$, obtained from direct diagrammatic calculations~\cite{Braun:2019wyx}. We apply it here to get the necessary two-loop $\Delta$ from the known two-loop light-light result~\cite{Braun:2016qlg}. 

The physical picture behind the light-to-heavy reduction from $\Delta_{qq}$ to $\Delta$ is the separation of collinear and soft divergences at the leading-power approximation. 
A massless quark can radiate collinear gluons that interact with either its partner quark or the finite lightlike Wilson line, hence relocating itself in the lightlike direction $n^\mu$ while inducing collinear singularities. By imposing the second quark as the heavy field of infinite mass, its interaction with either the soft gluon background or its light-quark counterpart cannot change its location, motivating us to discard any term that moves the second field in the original light-light results but maintaining the local constants and contributions that relocate only the remaining light field. This picture is explicitly verified at two-loop order in~\cite{Braun:2019wyx}  for the individual nontrivial exchange diagrams with both light and heavy quarks participating in the interaction. 

The structure of the flavor-nonsinglet Efremov–
Radyushkin–Brodsky–Lepage (ERBL) kernel and the LN kernel in the conformal representation at the Wilson-Fisher critical point provides another perspective. 
Taking the light-to-heavy limit at the operator level, the light-light $SL(2,R)$ Casimir operator for the ERBL kernel decouples into a product of the translation generator $S^{(h)}_-$ for the heavy quark and the special conformal generator for the surviving light quark $S_+^{(q)}$~\cite{Braun:2018fiz}. Since these two operators commute, the cusp term separates into a heavy-quark logarithm, absorbed into HQET matching up to a simple scheme-dependent constant, and $\cusp\ln[i\mu e^{2\gamma_E}\mathcal K]$, with quantum corrections included in $\mathcal K\sim S_+^{(q)}$. It can be shown that this reduction at the conformal operator level is equivalent to the position-space light-to-heavy reduction procedure described above. 
More detailed discussions of the reduction procedure are provided in the supplemental material. 

Employing the reduction, we obtain the following form of the $\Delta^{(2)}$ operator from $\Delta^{(2)}_{qq}$ in Ref.~\cite{Braun:2016qlg},
\begin{equation}
 [\Delta^{(2)} f](z)=X^{(2)} f(z)+\int_0^1 d\alpha\,w^{(2)}(\alpha)[f(z)-f(\bar \alpha z)].
 \label{eq:S2}
\end{equation}
Here $X^{(2)}$ and $w^{(2)}(\alpha)$ are the two-loop local constant and nonlocal contributions to the special conformal generator ${\cal K}$, respectively.  The Supplemental Material provides explicit expressions for both $X^{(2)}$ and $w^{(2)}(\alpha)$.

\textit{The local constant from Wilson-line geometry.}---The commutation relations leave $\Gamma_+$ in eq.~\eqref{eq:conformal} so far undetermined and it had to be obtained via diagrammatically calculation in~\cite{Braun:2019wyx}. In this Letter, we fix it through consistency requirements. Let us first consider the nonsinglet Dokshitzer–Gribov–Lipatov–Altarelli–Parisi (DGLAP) kernel~\cite{Altarelli:1977zs,Curci:1980uw,Moch:2004pa} at the large spin-$N$ limit,
\begin{equation}
 \gamma_{qq}(N)=2\cusp\ln(Ne^{\gamma_E})+C_{qq}
 +O(N^{-1}\ln^kN),
 \label{eq:large-spin}
\end{equation}
where $C_{qq}$ is the noncusp constant of our interest. 
It is clear from Eq.~\eqref{eq:conformal} that the eigenvalue spectrum of the LN kernel takes the form,
\begin{align}
    E_s=\Gamma_{\rm cusp}\ln(\mu se^{2\gamma_E})+\Gamma_+\, ,
\end{align}
with $s$ being the eigenvalue of the special conformal operator ${\cal K}$ in Eq.~\eqref{def:K} for the eigenfunction $Q_s(z)$ defined as $[{\cal K}Q_s](z) = -isQ_s(z)$. In other words, $\Gamma_+$ is the noncusp part of the eigenvalue spectrum of the LN kernel, which plays the same role as $C_{qq}$ for the DGLAP. This motivates us to expect that $\Gamma_+$ can be related to various known noncusp anomalous dimensions. 

\begin{figure}[!tbp]
 \centering
 \includegraphics[width=\columnwidth]{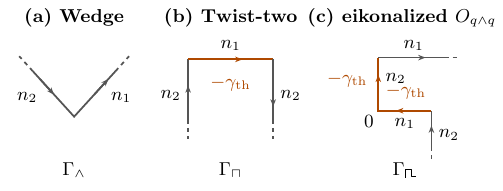}
 \caption{Soft Wilson loops for (a) the quark form factor, (b) the large-$N$ PDF, and (c) the matrix element of the $h\to\gamma\gamma$ soft-quark operator $O_{q\wedge q}$ with its external quarks eikonalized. Dashed lines indicate semi-infinite gauge links. The constants below each panel denote the corresponding noncusp soft anomalous dimensions.}
 \label{fig:geometry}
\end{figure}

To make progress, let us investigate the Drell--Yan (DY) production near partonic threshold where the cross section can be factorized as~\cite{Becher:2007ty},
\begin{align}
    \frac{d\sigma^{(q)}_{\rm LP}}{dQ^2} = \sigma_0^{(q)} H_{\rm DY}\times f_q\otimes f_{\bar q}\otimes S_{\rm DY}\, .
\end{align}
Here $\sigma_0^{(q)}$ is the scale-independent Born-level cross section and $Q^2$ is the invariant mass of the lepton pair. The DY hard function $H_{\rm DY}$ is the squared IR-subtracted quark form factor. $f_q$ and $f_{\bar q}$ are the parton distribution functions (PDFs) from the two incoming hadrons. The threshold DY soft function $S_{\rm DY}$ is defined as the squared matrix element of semi-infinite lightlike Wilson lines.  
In addition, the LP threshold approximation also dictates that only the cusp and endpoint effects, which correspond to the first two terms in the large $N$-expansion in eq.~\eqref{eq:large-spin}, contribute to the evolution of the two PDFs.  
This is because the large spin-$N$ expansion is equivalent to the $x\to1$ expansion for the splitting function $P_{q\leftarrow q}(x)$ in the momentum fraction space, so that radiated gluons carry only vanishingly small fraction of the incoming quark's longitudinal momentum as required by the DY threshold production. 
RG consistency then imposes the following conditions on the noncusp anomalous dimensions of the hard function ($4\gamma_{q}$) with $2\gamma_q$ being the anomalous dimension of the quark form factor, the PDFs ($-2C_{qq}$), and the threshold anomalous dimension ($-2\gamma_{\rm th}$), 
\begin{equation}
 4\gamma_q-2C_{qq}-2\gamma_{\rm th}=0.
 \label{eq:threshold-consistency}
\end{equation}
$\gamma_{\rm th}$ can also be expressed as a linear combination of the anomalous dimensions of various Wilson loops depicted in Fig.~\ref{fig:geometry}.  
The wedged Wilson loop in [Fig.~\ref{fig:geometry}(a)] with two semi-infinite lightlike Wilson lines meeting at a common point $0$ corresponds to the eikonalized light-quark form factor. It captures the soft-gluon contribution to the anomalous dimension of the quark form factor. 
The Wilson loop in [Fig.~\ref{fig:geometry}(b)] represents the eikonalization of the light-light operator $O_{qq}$ in Eq.~\eqref{def:Oqq}. It therefore only retains contributions generated by the soft-gluon interactions consistent with the large-$N$, or $x\to 1$ limit. 
Both configurations have noncusp anomalous dimensions, which we denote as $\Gamma_{\wedge}$ and $\Gamma_{\sqcap}$. The difference between $2\gamma_q$ and $\Gamma_{\wedge}$ is that $2\gamma_q$ contains extra contributions from hard-collinear radiations. At the same time $C_{qq} - \Gamma_{\sqcap}$ also removes all the soft-collinear contributions, leaving only the virtual hard-collinear corrections unique to $C_{qq}$. Since the hard-collinear corrections are universal as they are induced entirely by the light quarks, we conclude 
\begin{equation}
2\gamma_{q}-\Gamma_{\wedge} = C_{qq}-\Gamma_{\sqcap} = 2\gamma_{\rm hc}^{(q)}\, , 
  \label{eq:threshold-geometry}
\end{equation}
where $\gamma_{\rm hc}^{(q)}$ denotes the hard-collinear anomalous dimension from a single quark. 
The latter relation reveals that the full noncusp constant $C_{qq}$ can be reconstructed by combining contributions from the soft-collinear ($\Gamma_{\sqcap}$) and the hard-collinear regions. While the former only depends on the corresponding Wilson loop, the latter is local hence insensitive to the operator's geometric shape. Combining Eq.~\eqref{eq:threshold-geometry} and~\eqref{eq:threshold-consistency} yields,
\begin{align}\label{eq:wedge-sqcap}
    \Gamma_{\wedge}-\Gamma_{\sqcap} = \gamma_{\rm th}\, .
\end{align}

To finally determine $\Gamma_+$, we use the soft-quark operator $O_{q\wedge q}$ that defines the soft function entering $h\to\gamma\gamma$ at next-to-leading power (NLP). $O_{q\wedge q}$ is composed of two light quarks on different lightlike directions $(n_1,n_2)$ connected by two collinear gauge links at the origin~\cite{Liu:2019oav}. Its kernel is the sum of the LN kernels acting on the two quark fields, minus $2\gamma_h$, where $\gamma_h$ is the single-heavy-quark anomalous dimension~\cite{Becher:2009kw}. Its full constant term in the conformal representation is therefore $2(\Gamma_+-\gamma_h)$~\cite{Beneke:2024cpq}. Here, $2(\Gamma_+-\gamma_h)$ includes the localized hard-collinear contributions ($2\gamma_q-\Gamma_\wedge$) isolated in Eq.~\eqref{eq:threshold-geometry}, which are induced by the two light quarks. Removing them leaves us with $2(\Gamma_+-\gamma_h)-(2\gamma_q-\Gamma_\wedge)$, which is equal to the anomalous dimension $\Gamma_{\diphotoncontour}$ of the eikonalized $O_{q\wedge q}$ operator in the form of the Wilson loop shown in [Fig.~\ref{fig:geometry}(c)].

For closed lightlike polygon loops, the web analysis shows that the noncusp term is generated from regions where the gluon vertices approach the same lightlike Wilson line~\cite{Erdogan:2011yc}. The collinear divergence is hence induced by the gluon propagator in the schematic form of $\sim\tfrac{g_{\mu\nu}}{[-(x-y)^2+i0]^{1-\epsilon}}$ with $(x-y)^2\to0$. Further region analysis reveals that the noncusp anomalous dimensions for polygon-shaped Wilson loops composed of lightlike Wilson lines of different directions can be systematically constructed as linear combinations of simple noncusp anomalous dimensions, as demonstrated in Ref.~\cite{Falcioni:2019nxk}. In this framework, the two equivalent semi-infinite lightlike Wilson lines in [Fig.~\ref{fig:geometry}(a)] each contribute $\Gamma_{\wedge}/2$. In comparison to $\Gamma_\wedge$, to obtain $\Gamma_{\sqcap}$, we need to include extra contributions due to the presence of the finite link. 
Eq.~\eqref{eq:wedge-sqcap} helps us fix such addition contributions.  
More crucially, it indicates from [Fig.~\ref{fig:geometry}(b)] that $\Gamma_{\sqcap}$ can be constructed by attributing each lightlike semi-infinite (finite) Wilson line 
with a $\Gamma_\wedge/2$ ($-\gamma_{\rm th}$) term. 
Applying this factorization-based construction procedure to [Fig.~\ref{fig:geometry} (c)], we find,  
\begin{align}
    \Gamma_{\diphotoncontour} = \Gamma_{\wedge} - 2 \gamma_{\rm th} \stackrel{\text{Eq.~\eqref{eq:wedge-sqcap}}} = 2\Gamma_{\sqcap}-\Gamma_{\wedge}\, ,\label{eq:gamma-wiggle}
\end{align}
and together with $ \Gamma_{\diphotoncontour}  = 2(\Gamma_+-\gamma_h)-2\gamma_q+\Gamma_\wedge$, we obtain,
\begin{equation}
 \Gamma_+=C_{qq}+\gamma_h-\gamma_q=\gamma_q+\gamma_h-\gamma_{\rm th}\, .
 \label{eq:constant}
\end{equation}

Using available one- and two-loop results for $\Gamma_+$ computed through diagrammatic means~\cite{Braun:2014owa,Braun:2019wyx}, we verified the correctness of Eq.~\eqref{eq:constant}. 
We consequently determine the three-loop $\Gamma_+$ from the three-loop light-quark~\cite{Moch:2005id,Becher:2009qa}, heavy-quark~\cite{Grozin:2014hna,Hoang:2015vua}, and threshold~\cite{Moch:2005tm,Li:2014afw} anomalous dimensions.  The explicit expressions are collected in the Supplemental Material.

\textit{Three-loop kernel.}---We subsequently derive the position-space representation from Eq.~\eqref{eq:conformal} by Mellin transform. The kernel takes the form in Eq.~\eqref{eq:kernel} as expected, with all constants expanded to $a_s^3$. The non-local corrections are incorporated in the function $h(\alpha)$, which is now known to the next-to-next-to-leading order. The complete expressions for both $h(\alpha)$ and $\gamma_+$ defined in Eq.~\eqref{eq:kernel}, together with $\cusp$ up to ${\cal O}(a_s^3)$, are provided in the Supplemental Material, where the LN kernel in momentum space is also collected. The new function $h^{(2)}(\alpha)$ contains harmonic polylogarithms with transcendental weights at most four, with indices only $0$ and $1$~\cite{Remiddi:1999ew}.

\textit{Consistency checks.}---At leading power, the radiative leptonic $B$-meson decay form factors factorize as~\cite{Descotes-Genon:2002crx,Lunghi:2002ju,Bosch:2003fc}
\begin{eqnarray}
 F_{V,A}^{\rm (LP)}(E_\gamma) &=&\frac{Q_u \, m_B \, f_B }{2 \, E_\gamma} \, H(E_\gamma, \mu) \, 
 \nonumber \\
&&  \times\int_0^\infty\frac{d\omega}{\omega} \, J_\gamma(E_\gamma \, \omega, \mu) \, \phi_B^{+}(\omega, \mu).
 \label{eq:lp-factorization}
\end{eqnarray}
Here $E_\gamma$ is the photon energy in the meson rest frame, $Q_u=2/3$, $m_B$ is the meson mass, with $H$ and $J_\gamma$ being the hard and photonic radiative jet functions. After implementing ultraviolet renormalization and infrared subtraction, $J_\gamma=Z_J\otimes J_\gamma^{\rm bare}$. Scale independence fixes $Z_J$ from the LN kernel together with the hard  anomalous dimensions. See Supplemental Material for more detailed discussions about the jet function renormalization constant and anomalous dimension. We have verified that the thus derived $Z_J$ to ${\cal O}(a_s^3)$ cancels all $\epsilon$-poles of the independently calculated three-loop bare jet function; see our companion paper for details~\cite{Huang:3Ljet}. This offers highly nontrivial mutual validation to both our three-loop LN kernel and the three-loop jet function. 

As an additional consistency check, we apply the light-to-heavy reduction procedure to the newly available complete three-loop position-space ERBL kernel~\cite{Chen:3LERBL} and find full agreement after removing the cusp logarithm term and adjusting for the local term using $\gamma_q$ and $\gamma_h$. We provide some details of this comparison in the Supplemental Material.

\textit{Conclusions.}---We have bootstrapped the complete three-loop LN kernel using conformal technique and known anomalous dimensions, without performing any diagrammatic calculations. The light-to-heavy reduction principle provides us with the two-loop conformal anomaly needed for constructing the nonlocal part of the LN kernel at three loops. For the constant part, the comparison between various Wilson-line geometries gives an all-order physical argument for $\Gamma_+=\gamma_q+\gamma_h-\gamma_{\rm th}$. Our three-loop LN kernel is consistent with the pole structures of the independently calculated radiative-jet function and, in a subsequent check, the light-to-heavy limit of the companion ERBL kernel.  Our result enables the NNLL evolution of the leading-twist $B$-meson LCDA and is also directly applicable to the evolution of the soft function appearing in the NLP $h\to\gamma\gamma$ decay.

\vspace{0.5 cm}

\paragraph{\textcolor{black}{Note Added}:} The main results of this Letter  have  been reported at an  international  workshop 
{\it Precision Across Scales: from QCD to the cosmos} [{\tt https://indico.ph.tum.de/event/8036/}]
and   at a national   workshop on cutting-edge research  of theoretical physics
[{\tt https://indico.ihep.ac.cn/event/30082/}].
\vspace{0.2 cm}

\paragraph{\textcolor{black}{Acknowledgments}}

Y.J. is grateful for the support from  the National Natural Science Foundation of China with Grant No.12535006, from the Guangdong Basic and Applied Basic Research Foundation under Grant No. 2026A1515011306, and from  the University Development Fund of the Chinese University of Hong Kong, Shenzhen, under the Grant No. UDF01003869. B.X.S. acknowledges support from the National Natural Science Foundation of China under Grant No. 12325503. X.W. is supported by the University Development Fund of The Chinese University of Hong Kong, Shenzhen, under  Grant No. UDF01003912 and by the National Natural Science Foundation of China with Grant No. 12535006. Y.M.W. acknowledges support from the  National Natural Science Foundation of China with Grants No. 12475097 and No. 12535006, from  the Natural Science Foundation of Tianjin with Grant No. 25JCZDJC01190, and from the Fundamental Research Funds for the Central Universities with Grant No. 63261180.

\bibliographystyle{apsrev4-2}
\bibliography{references}

@article{Grozin:1996pq,
    author = "Grozin, A. G. and Neubert, M.",
    title = "{Asymptotics of heavy meson form-factors}",
    eprint = "hep-ph/9607366",
    archivePrefix = "arXiv",
    reportNumber = "BUDKER-INP-1996-45, BINP-96-45, CERN-TH-96-144",
    doi = "10.1103/PhysRevD.55.272",
    journal = "Phys. Rev. D",
    volume = "55",
    pages = "272--290",
    year = "1997"
}

@article{Beneke:1999br,
    author = "Beneke, M. and Buchalla, G. and Neubert, M. and Sachrajda, Christopher T.",
    title = "{QCD factorization for B ---{\ensuremath{>}} pi pi decays: Strong phases and CP violation in the heavy quark limit}",
    eprint = "hep-ph/9905312",
    archivePrefix = "arXiv",
    reportNumber = "SLAC-PUB-8146, CERN-TH-99-126, SHEP-99-04",
    doi = "10.1103/PhysRevLett.83.1914",
    journal = "Phys. Rev. Lett.",
    volume = "83",
    pages = "1914--1917",
    year = "1999"
}

@article{Beneke:2001at,
    author = "Beneke, M. and Feldmann, T. and Seidel, D.",
    title = "{Systematic approach to exclusive $B \to  V l^+ l^-$, $V \gamma$ decays}",
    eprint = "hep-ph/0106067",
    archivePrefix = "arXiv",
    reportNumber = "PITHA-01-05",
    doi = "10.1016/S0550-3213(01)00366-2",
    journal = "Nucl. Phys. B",
    volume = "612",
    pages = "25--58",
    year = "2001"
}

@article{Beneke:2018wjp,
    author = "Beneke, M. and Braun, V. M. and Ji, Yao and Wei, Yan-Bing",
    title = "{Radiative leptonic decay $B\to \gamma \ell \nu_\ell$ with subleading power corrections}",
    eprint = "1804.04962",
    archivePrefix = "arXiv",
    primaryClass = "hep-ph",
    reportNumber = "TUM-HEP-1135-18",
    doi = "10.1007/JHEP07(2018)154",
    journal = "JHEP",
    volume = "2018",
    number = "07",
    pages = "154",
    year = "2018"
}

@article{Liu:2020ydl,
    author = "Liu, Ze Long and Neubert, Matthias",
    title = "{Two-Loop Radiative Jet Function for Exclusive $B$-Meson and Higgs Decays}",
    eprint = "2003.03393",
    archivePrefix = "arXiv",
    primaryClass = "hep-ph",
    reportNumber = "MITP/20-011",
    doi = "10.1007/JHEP06(2020)060",
    journal = "JHEP",
    volume = "2020",
    number = "06",
    pages = "060",
    year = "2020"
}

@article{Liu:2020eqe,
    author = "Liu, Ze Long and Mecaj, Bianka and Neubert, Matthias and Wang, Xing and Fleming, Sean",
    title = "{Renormalization and Scale Evolution of the Soft-Quark Soft Function}",
    eprint = "2005.03013",
    archivePrefix = "arXiv",
    primaryClass = "hep-ph",
    reportNumber = "MITP/20-023",
    doi = "10.1007/JHEP07(2020)104",
    journal = "JHEP",
    volume = "2020",
    number = "07",
    pages = "104",
    year = "2020"
}

@article{Beneke:2024cpq,
    author = "Beneke, Martin and Ji, Yao and Wang, Xing",
    title = "{Renormalization of the next-to-leading-power {\ensuremath{\gamma}}{\ensuremath{\gamma}} {\textrightarrow} h and gg {\textrightarrow} h soft quark functions}",
    eprint = "2403.17738",
    archivePrefix = "arXiv",
    primaryClass = "hep-ph",
    reportNumber = "TUM-HEP-1501/24",
    doi = "10.1007/JHEP05(2024)246",
    journal = "JHEP",
    volume = "2024",
    number = "05",
    pages = "246",
    year = "2024"
}

@article{Lange:2003ff,
    author = "Lange, Bjorn O. and Neubert, Matthias",
    title = "{Renormalization group evolution of the B meson light cone distribution amplitude}",
    eprint = "hep-ph/0303082",
    archivePrefix = "arXiv",
    reportNumber = "CLNS-03-1822",
    doi = "10.1103/PhysRevLett.91.102001",
    journal = "Phys. Rev. Lett.",
    volume = "91",
    pages = "102001",
    year = "2003"
}

@article{Bell:2013tfa,
    author = "Bell, Guido and Feldmann, Thorsten and Wang, Yu-Ming and Yip, Matthew W Y",
    title = "{Light-Cone Distribution Amplitudes for Heavy-Quark Hadrons}",
    eprint = "1308.6114",
    archivePrefix = "arXiv",
    primaryClass = "hep-ph",
    reportNumber = "OUTP-13-15P, SI-HEP-2013-05, QFET-2013-05, TTK-13-19, SFB-CPP-13-60, IPPP-13-55, DCPT-13-110, TUM-HEP-899-13",
    doi = "10.1007/JHEP11(2013)191",
    journal = "JHEP",
    volume = "2013",
    number = "11",
    pages = "191",
    year = "2013"
}

@article{Braun:2014owa,
    author = "Braun, V. M. and Manashov, A. N.",
    title = "{Conformal symmetry of the Lange-Neubert evolution equation}",
    eprint = "1402.5822",
    archivePrefix = "arXiv",
    primaryClass = "hep-ph",
    doi = "10.1016/j.physletb.2014.02.051",
    journal = "Phys. Lett. B",
    volume = "731",
    pages = "316--319",
    year = "2014"
}

@article{Braun:2019wyx,
    author = "Braun, V. M. and Ji, Yao and Manashov, A. N.",
    title = "{Two-loop evolution equation for the B-meson distribution amplitude}",
    eprint = "1905.04498",
    archivePrefix = "arXiv",
    primaryClass = "hep-ph",
    reportNumber = "DESY-19-080",
    doi = "10.3204/PUBDB-2019-02451",
    journal = "Phys. Rev. D",
    volume = "100",
    number = "1",
    pages = "014023",
    year = "2019"
}

@article{Braun:2018mxm,
    author = "Braun, V. M. and Manashov, A. N. and Moch, S. O. and Strohmaier, M.",
    title = "{Conformal symmetry of QCD in $d$-dimensions}",
    eprint = "1810.04993",
    archivePrefix = "arXiv",
    primaryClass = "hep-th",
    reportNumber = "DESY-18-175, DESY 18-175",
    doi = "10.1016/j.physletb.2019.04.027",
    journal = "Phys. Lett. B",
    volume = "793",
    pages = "78--84",
    year = "2019"
}

@article{Braun:2016qlg,
    author = "Braun, V. M. and Manashov, A. N. and Moch, S. and Strohmaier, M.",
    title = "{Two-loop conformal generators for leading-twist operators in QCD}",
    eprint = "1601.05937",
    archivePrefix = "arXiv",
    primaryClass = "hep-ph",
    reportNumber = "DESY-16-012",
    doi = "10.1007/JHEP03(2016)142",
    journal = "JHEP",
    volume = "2016",
    number = "03",
    pages = "142",
    year = "2016"
}

@article{Braun:2017cih,
    author = "Braun, V. M. and Manashov, A. N. and Moch, S. and Strohmaier, M.",
    title = "{Three-loop evolution equation for flavor-nonsinglet operators in off-forward kinematics}",
    eprint = "1703.09532",
    archivePrefix = "arXiv",
    primaryClass = "hep-ph",
    reportNumber = "DESY-17-045",
    doi = "10.1007/JHEP06(2017)037",
    journal = "JHEP",
    volume = "2017",
    number = "06",
    pages = "037",
    year = "2017"
}

@article{Braun:2020yib,
    author = "Braun, V. M. and Manashov, A. N. and Moch, S. and Schoenleber, J.",
    title = "{Two-loop coefficient function for DVCS: vector contributions}",
    eprint = "2007.06348",
    archivePrefix = "arXiv",
    primaryClass = "hep-ph",
    reportNumber = "DESY-20-116, DESY 20-116",
    doi = "10.1007/JHEP09(2020)117",
    journal = "JHEP",
    volume = "2020",
    number = "09",
    pages = "117",
    year = "2020",
    note = "[Erratum: JHEP 02, 115 (2022)]"
}

@article{Braun:2018fiz,
    author = "Braun, Vladimir M. and Ji, Yao and Manashov, Alexander N.",
    title = "{Integrability in heavy quark effective theory}",
    eprint = "1804.06289",
    archivePrefix = "arXiv",
    primaryClass = "hep-th",
    reportNumber = "DESY-18-056",
    doi = "10.1007/JHEP06(2018)017",
    journal = "JHEP",
    volume = "2018",
    number = "06",
    pages = "017",
    year = "2018"
}

@article{Chetyrkin:2003vi,
    author = "Chetyrkin, K. G. and Grozin, A. G.",
    title = "{Three loop anomalous dimension of the heavy light quark current in HQET}",
    eprint = "hep-ph/0303113",
    archivePrefix = "arXiv",
    reportNumber = "TTP03-10, SFB-CPP-03-05, THEP-03-05",
    doi = "10.1016/S0550-3213(03)00490-5",
    journal = "Nucl. Phys. B",
    volume = "666",
    pages = "289--302",
    year = "2003"
}

@article{Moch:2004pa,
    author = "Moch, S. and Vermaseren, J. A. M. and Vogt, A.",
    title = "{The Three loop splitting functions in QCD: The Nonsinglet case}",
    eprint = "hep-ph/0403192",
    archivePrefix = "arXiv",
    reportNumber = "DESY-04-047, SFB-CPP-04-09, NIKHEF-04-001",
    doi = "10.1016/j.nuclphysb.2004.03.030",
    journal = "Nucl. Phys. B",
    volume = "688",
    pages = "101--134",
    year = "2004"
}

@article{Moch:2005id,
    author = "Moch, S. and Vermaseren, J. A. M. and Vogt, A.",
    title = "{The Quark form-factor at higher orders}",
    eprint = "hep-ph/0507039",
    archivePrefix = "arXiv",
    reportNumber = "SFB-CPP-05-26, DCPT-05-66, IPP-05-33, NIKHEF-05-011, DESY-05-106",
    doi = "10.1088/1126-6708/2005/08/049",
    journal = "JHEP",
    volume = "2005",
    number = "08",
    pages = "049",
    year = "2005"
}

@article{Hoang:2015vua,
    author = "Hoang, Andre H. and Pathak, Aditya and Pietrulewicz, Piotr and Stewart, Iain W.",
    title = "{Hard Matching for Boosted Tops at Two Loops}",
    eprint = "1508.04137",
    archivePrefix = "arXiv",
    primaryClass = "hep-ph",
    reportNumber = "UWTHPH-2015-18, MIT-CTP-4693, DESY-15-139",
    doi = "10.1007/JHEP12(2015)059",
    journal = "JHEP",
    volume = "2015",
    number = "12",
    pages = "059",
    year = "2015"
}

@article{Li:2014afw,
    author = "Li, Ye and von Manteuffel, Andreas and Schabinger, Robert M. and Zhu, Hua Xing",
    title = "{Soft-virtual corrections to Higgs production at N$^3$LO}",
    eprint = "1412.2771",
    archivePrefix = "arXiv",
    primaryClass = "hep-ph",
    reportNumber = "SLAC-PUB-16169, MITP-14-095",
    doi = "10.1103/PhysRevD.91.036008",
    journal = "Phys. Rev. D",
    volume = "91",
    pages = "036008",
    year = "2015"
}

@article{Braun:2003wx,
    author = "Braun, V. M. and Ivanov, D. Yu. and Korchemsky, G. P.",
    title = "{The B meson distribution amplitude in QCD}",
    eprint = "hep-ph/0309330",
    archivePrefix = "arXiv",
    reportNumber = "LPT-ORSAY-03-63",
    doi = "10.1103/PhysRevD.69.034014",
    journal = "Phys. Rev. D",
    volume = "69",
    pages = "034014",
    year = "2004"
}

@article{Falcioni:2019nxk,
    author = "Falcioni, Giulio and Gardi, Einan and Milloy, Calum",
    title = "{Relating amplitude and PDF factorisation through Wilson-line geometries}",
    eprint = "1909.00697",
    archivePrefix = "arXiv",
    primaryClass = "hep-ph",
    doi = "10.1007/JHEP11(2019)100",
    journal = "JHEP",
    volume = "2019",
    number = "11",
    pages = "100",
    year = "2019"
}

@article{Becher:2009kw,
    author = "Becher, Thomas and Neubert, Matthias",
    title = "{Infrared singularities of QCD amplitudes with massive partons}",
    eprint = "0904.1021",
    archivePrefix = "arXiv",
    primaryClass = "hep-ph",
    reportNumber = "FERMILAB-PUB-09-099-T, MZ-TH-09-11",
    doi = "10.1103/PhysRevD.79.125004",
    journal = "Phys. Rev. D",
    volume = "79",
    pages = "125004",
    year = "2009",
    note = "[Erratum: Phys.Rev.D 80, 109901 (2009)]"
}

@article{Altarelli:1977zs,
    author = "Altarelli, Guido and Parisi, G.",
    title = "{Asymptotic Freedom in Parton Language}",
    reportNumber = "LPTENS-77-6",
    doi = "10.1016/0550-3213(77)90384-4",
    journal = "Nucl. Phys. B",
    volume = "126",
    pages = "298--318",
    year = "1977"
}

@article{Curci:1980uw,
    author = "Curci, G. and Furmanski, W. and Petronzio, R.",
    title = "{Evolution of Parton Densities Beyond Leading Order: The Nonsinglet Case}",
    reportNumber = "CERN-TH-2815",
    doi = "10.1016/0550-3213(80)90003-6",
    journal = "Nucl. Phys. B",
    volume = "175",
    pages = "27--92",
    year = "1980"
}

@article{Becher:2009qa,
    author = "Becher, Thomas and Neubert, Matthias",
    title = "{On the Structure of Infrared Singularities of Gauge-Theory Amplitudes}",
    eprint = "0903.1126",
    archivePrefix = "arXiv",
    primaryClass = "hep-ph",
    reportNumber = "FERMILAB-PUB-09-061-T, MZ-TH-09-05",
    doi = "10.1088/1126-6708/2009/06/081",
    journal = "JHEP",
    volume = "2009",
    number = "06",
    pages = "081",
    year = "2009",
    note = "[Erratum: JHEP 11, 024 (2013)]"
}

@article{Korchemsky:1987wg,
    author = "Korchemsky, G. P. and Radyushkin, A. V.",
    title = "{Renormalization of the Wilson Loops Beyond the Leading Order}",
    doi = "10.1016/0550-3213(87)90277-X",
    journal = "Nucl. Phys. B",
    volume = "283",
    pages = "342--364",
    year = "1987"
}

@article{Grozin:2014hna,
    author = "Grozin, Andrey and Henn, Johannes M. and Korchemsky, Gregory P. and Marquard, Peter",
    title = "{Three Loop Cusp Anomalous Dimension in QCD}",
    eprint = "1409.0023",
    archivePrefix = "arXiv",
    primaryClass = "hep-ph",
    reportNumber = "IPHT-T14-111, DESY-14-148, SFB-CPP-14-64, LPN14-104",
    doi = "10.1103/PhysRevLett.114.062006",
    journal = "Phys. Rev. Lett.",
    volume = "114",
    number = "6",
    pages = "062006",
    year = "2015"
}

@article{Becher:2007ty,
    author = "Becher, Thomas and Neubert, Matthias and Xu, Gang",
    title = "{Dynamical Threshold Enhancement and Resummation in Drell-Yan Production}",
    eprint = "0710.0680",
    archivePrefix = "arXiv",
    primaryClass = "hep-ph",
    reportNumber = "CLNS-07-2009, FERMILAB-PUB-07-507-T, MZ-TH-07-16",
    doi = "10.1088/1126-6708/2008/07/030",
    journal = "JHEP",
    volume = "2008",
    number = "07",
    pages = "030",
    year = "2008"
}

@article{Moch:2005tm,
    author = "Moch, S. and Vermaseren, J. A. M. and Vogt, A.",
    title = "{Three-loop results for quark and gluon form-factors}",
    eprint = "hep-ph/0508055",
    archivePrefix = "arXiv",
    reportNumber = "DESY-05-138, SFT-CPP-05-34, NIKHEF-05-013, DCPT-05-88, IPPP-05-44",
    doi = "10.1016/j.physletb.2005.08.067",
    journal = "Phys. Lett. B",
    volume = "625",
    pages = "245--252",
    year = "2005"
}

@article{ATLAS:2022tnm,
    author = "Aad, Georges and others",
    collaboration = "ATLAS",
    title = "{Measurement of the properties of Higgs boson production at $\sqrt{s} = 13$ TeV in the $H\to\gamma\gamma$ channel using $139$ fb$^{-1}$ of $pp$ collision data with the ATLAS experiment}",
    eprint = "2207.00348",
    archivePrefix = "arXiv",
    primaryClass = "hep-ex",
    reportNumber = "CERN-EP-2022-094",
    doi = "10.1007/JHEP07(2023)088",
    journal = "JHEP",
    volume = "2023",
    number = "07",
    pages = "088",
    year = "2023"
}

@article{Belitsky:1999gu,
    author = "Belitsky, Andrei V. and Mueller, Dieter and Freund, A.",
    title = "{Reconstruction of nonforward evolution kernels}",
    eprint = "hep-ph/9904477",
    archivePrefix = "arXiv",
    doi = "10.1016/S0370-2693(99)00837-0",
    journal = "Phys. Lett. B",
    volume = "461",
    pages = "270--279",
    year = "1999"
}

@article{Beneke:2000wa,
    author = "Beneke, M. and Feldmann, T.",
    title = "{Symmetry breaking corrections to heavy to light B meson form-factors at large recoil}",
    eprint = "hep-ph/0008255",
    archivePrefix = "arXiv",
    reportNumber = "PITHA-00-20",
    doi = "10.1016/S0550-3213(00)00585-X",
    journal = "Nucl. Phys. B",
    volume = "592",
    pages = "3--34",
    year = "2001"
}

@article{Beneke:2011nf,
    author = "Beneke, M. and Rohrwild, J.",
    title = "{B meson distribution amplitude from B --{\ensuremath{>}} {\textbackslash}gamma l {\textbackslash}nu}",
    eprint = "1110.3228",
    archivePrefix = "arXiv",
    primaryClass = "hep-ph",
    reportNumber = "TTK-11-51, SFB-CPP-11-56",
    doi = "10.1140/epjc/s10052-011-1818-8",
    journal = "Eur. Phys. J. C",
    volume = "71",
    pages = "1818",
    year = "2011"
}

@article{Bosch:2003fc,
    author = "Bosch, S. W. and Hill, R. J. and Lange, B. O. and Neubert, M.",
    title = "{Factorization and Sudakov resummation in leptonic radiative B decay}",
    eprint = "hep-ph/0301123",
    archivePrefix = "arXiv",
    reportNumber = "SLAC-PUB-9625, CLNS-03-1817",
    doi = "10.1103/PhysRevD.67.094014",
    journal = "Phys. Rev. D",
    volume = "67",
    pages = "094014",
    year = "2003"
}

@article{Braun:2013tva,
    author = "Braun, V. M. and Manashov, A. N.",
    title = "{Evolution equations beyond one loop from conformal symmetry}",
    eprint = "1306.5644",
    archivePrefix = "arXiv",
    primaryClass = "hep-th",
    doi = "10.1140/epjc/s10052-013-2544-1",
    journal = "Eur. Phys. J. C",
    volume = "73",
    pages = "2544",
    year = "2013"
}

@article{Braun:2014vba,
    author = "Braun, V. M. and Manashov, A. N.",
    title = "{Two-loop evolution equations for light-ray operators}",
    eprint = "1404.0863",
    archivePrefix = "arXiv",
    primaryClass = "hep-ph",
    doi = "10.1016/j.physletb.2014.05.037",
    journal = "Phys. Lett. B",
    volume = "734",
    pages = "137--143",
    year = "2014"
}

@article{CMS:2021kom,
    author = "Sirunyan, Albert M and others",
    collaboration = "CMS",
    title = "{Measurements of Higgs boson production cross sections and couplings in the diphoton decay channel at $ \sqrt{\mathrm{s}} $ = 13 TeV}",
    eprint = "2103.06956",
    archivePrefix = "arXiv",
    primaryClass = "hep-ex",
    reportNumber = "CMS-HIG-19-015, CERN-EP-2021-038",
    doi = "10.1007/JHEP07(2021)027",
    journal = "JHEP",
    volume = "2021",
    number = "07",
    pages = "027",
    year = "2021"
}

@article{DeFazio:2005dx,
    author = "De Fazio, Fulvia and Feldmann, Thorsten and Hurth, Tobias",
    title = "{Light-cone sum rules in soft-collinear effective theory}",
    eprint = "hep-ph/0504088",
    archivePrefix = "arXiv",
    reportNumber = "CERN-PH-TH-2005-053, SLAC-PUB-11082, SI-HEP-2005-03, BARI-TH-05-509",
    doi = "10.1016/j.nuclphysb.2005.09.047",
    journal = "Nucl. Phys. B",
    volume = "733",
    pages = "1--30",
    year = "2006",
    note = "[Erratum: Nucl.Phys.B 800, 405 (2008)]"
}

@article{Descotes-Genon:2002crx,
    author = "Descotes-Genon, S. and Sachrajda, C. T.",
    title = "{Factorization, the light cone distribution amplitude of the B meson and the radiative decay B ---{\ensuremath{>}} gamma l nu(l)}",
    eprint = "hep-ph/0209216",
    archivePrefix = "arXiv",
    reportNumber = "CERN-TH-2002-228, SHEP-02-16",
    doi = "10.1016/S0550-3213(02)01066-0",
    journal = "Nucl. Phys. B",
    volume = "650",
    pages = "356--390",
    year = "2003"
}

@article{Efremov:1979qk,
    author = "Efremov, A. V. and Radyushkin, A. V.",
    title = "{Factorization and Asymptotical Behavior of Pion Form-Factor in QCD}",
    reportNumber = "JINR-P2-12900",
    doi = "10.1016/0370-2693(80)90869-2",
    journal = "Phys. Lett. B",
    volume = "94",
    pages = "245--250",
    year = "1980"
}

@article{Korchemsky:1999qb,
    author = "Korchemsky, Gregory P. and Pirjol, Dan and Yan, Tung-Mow",
    title = "{Radiative leptonic decays of B mesons in QCD}",
    eprint = "hep-ph/9911427",
    archivePrefix = "arXiv",
    reportNumber = "CLNS-99-1644, LPT-ORSAY-99-93, UCSD-PTH-99-17",
    doi = "10.1103/PhysRevD.61.114510",
    journal = "Phys. Rev. D",
    volume = "61",
    pages = "114510",
    year = "2000"
}

@article{Kotsky:1997rq,
    author = "Kotsky, M. I. and Yakovlev, Oleg I.",
    title = "{On the resummation of double logarithms in the process Higgs ---{\ensuremath{>}} gamma gamma}",
    eprint = "hep-ph/9708485",
    archivePrefix = "arXiv",
    reportNumber = "BUDKER-INP-1997-21, WUE-ITP-97-029",
    doi = "10.1016/S0370-2693(97)01260-4",
    journal = "Phys. Lett. B",
    volume = "418",
    pages = "335--344",
    year = "1998"
}

@article{Lepage:1979zb,
    author = "Lepage, G. Peter and Brodsky, Stanley J.",
    title = "{Exclusive Processes in Quantum Chromodynamics: Evolution Equations for Hadronic Wave Functions and the Form-Factors of Mesons}",
    reportNumber = "SLAC-PUB-2343",
    doi = "10.1016/0370-2693(79)90554-9",
    journal = "Phys. Lett. B",
    volume = "87",
    pages = "359--365",
    year = "1979"
}

@article{Liu:2019oav,
    author = "Liu, Ze Long and Neubert, Matthias",
    title = "{Factorization at subleading power and endpoint-divergent convolutions in $h\to\gamma\gamma$ decay}",
    eprint = "1912.08818",
    archivePrefix = "arXiv",
    primaryClass = "hep-ph",
    reportNumber = "MITP/19-078",
    doi = "10.1007/JHEP04(2020)033",
    journal = "JHEP",
    volume = "2020",
    number = "04",
    pages = "033",
    year = "2020"
}

@article{Liu:2020tzd,
    author = "Liu, Ze Long and Mecaj, Bianka and Neubert, Matthias and Wang, Xing",
    title = "{Factorization at subleading power, Sudakov resummation, and endpoint divergences in soft-collinear effective theory}",
    eprint = "2009.04456",
    archivePrefix = "arXiv",
    primaryClass = "hep-ph",
    reportNumber = "MITP/20-050, ZU-TH 31/20",
    doi = "10.1103/PhysRevD.104.014004",
    journal = "Phys. Rev. D",
    volume = "104",
    number = "1",
    pages = "014004",
    year = "2021"
}

@article{Liu:2020wbn,
    author = "Liu, Ze Long and Mecaj, Bianka and Neubert, Matthias and Wang, Xing",
    title = "{Factorization at subleading power and endpoint divergences in $h\to\gamma\gamma$ decay. Part II. Renormalization and scale evolution}",
    eprint = "2009.06779",
    archivePrefix = "arXiv",
    primaryClass = "hep-ph",
    reportNumber = "MITP/20-047",
    doi = "10.1007/JHEP01(2021)077",
    journal = "JHEP",
    volume = "2021",
    number = "01",
    pages = "077",
    year = "2021"
}

@article{Lunghi:2002ju,
    author = "Lunghi, Enrico and Pirjol, Dan and Wyler, Daniel",
    title = "{Factorization in leptonic radiative B $\to \gamma$e{\ensuremath{\nu}} decays}",
    eprint = "hep-ph/0210091",
    archivePrefix = "arXiv",
    reportNumber = "ZU-TH-17-02",
    doi = "10.1016/S0550-3213(02)01032-5",
    journal = "Nucl. Phys. B",
    volume = "649",
    pages = "349--364",
    year = "2003"
}

@article{Mueller:1993hg,
    author = "Mueller, Dieter",
    title = "{Conformal constraints and the evolution of the nonsinglet meson distribution amplitude}",
    reportNumber = "SLAC-PUB-6247",
    doi = "10.1103/PhysRevD.49.2525",
    journal = "Phys. Rev. D",
    volume = "49",
    pages = "2525--2535",
    year = "1994"
}

@article{Remiddi:1999ew,
    author = "Remiddi, E. and Vermaseren, J. A. M.",
    title = "{Harmonic polylogarithms}",
    eprint = "hep-ph/9905237",
    archivePrefix = "arXiv",
    reportNumber = "NIKHEF-99-005, TTP-99-08",
    doi = "10.1142/S0217751X00000367",
    journal = "Int. J. Mod. Phys. A",
    volume = "15",
    pages = "725--754",
    year = "2000"
}

@misc{Chen:3LERBL,
    author = "Chen, Long-Bin and Ji, Yao and Shi, Bo-Xuan and Wang, Yu-Ming",
    note = "{to appear}"
}

@misc{Huang:3Ljet,
    author = "Huang, Yong-Kang and Ji, Yao and Shi, Bo-Xuan and Wang, Xing and Wang, Yu-Ming and Wei, Yuxiang",
    note = "{to appear}"
}

@article{Braun:2003rp,
    author = {Braun, V. M. and Korchemsky, G. P. and M{\"u}ller, Dieter},
    title = "{The Uses of conformal symmetry in QCD}",
    eprint = "hep-ph/0306057",
    archivePrefix = "arXiv",
    doi = "10.1016/S0146-6410(03)90004-4",
    journal = "Prog. Part. Nucl. Phys.",
    volume = "51",
    pages = "311--398",
    year = "2003"
}

@article{Grozin:2023dlk,
    author = "Grozin, Andrey",
    title = "{Anomalous dimension of the heavy-light quark current in HQET up to four loops}",
    eprint = "2311.09894",
    archivePrefix = "arXiv",
    primaryClass = "hep-ph",
    doi = "10.1007/JHEP02(2024)198",
    journal = "JHEP",
    volume = "02",
    pages = "198",
    year = "2024"
}

@article{Erdogan:2011yc,
    author = "Erdo{\u{g}}an, Ozan and Sterman, George",
    title = "{Gauge Theory Webs and Surfaces}",
    eprint = "1112.4564",
    archivePrefix = "arXiv",
    primaryClass = "hep-th",
    reportNumber = "YITP-SB-11-33",
    doi = "10.1103/PhysRevD.91.016003",
    journal = "Phys. Rev. D",
    volume = "91",
    number = "1",
    pages = "016003",
    year = "2015"
}

@article{Braun:2015pha,
    author = "Braun, V. M. and Manashov, A. N. and Offen, N.",
    title = "{Evolution equation for the higher-twist B-meson distribution amplitude}",
    eprint = "1507.03445",
    archivePrefix = "arXiv",
    primaryClass = "hep-ph",
    reportNumber = "DESY-15-108",
    doi = "10.1103/PhysRevD.92.074044",
    journal = "Phys. Rev. D",
    volume = "92",
    number = "7",
    pages = "074044",
    year = "2015"
}

@article{Braun:2017liq,
    author = "Braun, V. M. and Ji, Yao and Manashov, A. N.",
    title = "{Higher-twist B-meson Distribution Amplitudes in HQET}",
    eprint = "1703.02446",
    archivePrefix = "arXiv",
    primaryClass = "hep-ph",
    reportNumber = "DESY-17-037",
    doi = "10.1007/JHEP05(2017)022",
    journal = "JHEP",
    volume = "05",
    pages = "022",
    year = "2017"
}

@article{Shen:2020hfq,
    author = "Shen, Yue-Long and Wang, Yu-Ming and Wei, Yan-Bing",
    title = "{Precision calculations of the double radiative bottom-meson decays in soft-collinear effective theory}",
    eprint = "2009.02723",
    archivePrefix = "arXiv",
    primaryClass = "hep-ph",
    doi = "10.1007/JHEP12(2020)169",
    journal = "JHEP",
    volume = "12",
    pages = "169",
    year = "2020"
}

@article{Beneke:2020fot,
    author = "Beneke, Martin and Bobeth, Christoph and Wang, Yu-Ming",
    title = "{$B_{d,s}\to\gamma\ell\bar{\ell}$ decay with an energetic photon}",
    eprint = "2008.12494",
    archivePrefix = "arXiv",
    primaryClass = "hep-ph",
    reportNumber = "TUM-HEP-1280/20",
    doi = "10.1007/JHEP12(2020)148",
    journal = "JHEP",
    volume = "12",
    pages = "148",
    year = "2020"
}

@article{Wang:2021yrr,
    author = "Wang, Chao and Wang, Yu-Ming and Wei, Yan-Bing",
    title = "{QCD factorization for the four-body leptonic B-meson decays}",
    eprint = "2111.11811",
    archivePrefix = "arXiv",
    primaryClass = "hep-ph",
    reportNumber = "TUM-HEP-1332/21",
    doi = "10.1007/JHEP02(2022)141",
    journal = "JHEP",
    volume = "02",
    pages = "141",
    year = "2022"
}

@article{Beneke:2021rjf,
    author = {Beneke, Martin and B{\"o}er, Philipp and Rigatos, Panagiotis and Vos, Kimberley Keri},
    title = "{QCD factorization of the four-lepton decay $B^-\rightarrow \ell \bar{\nu }_\ell \ell ^{(\prime )} \bar{\ell }^{(\prime )}$}",
    eprint = "2102.10060",
    archivePrefix = "arXiv",
    primaryClass = "hep-ph",
    reportNumber = "TUM-HEP-1317/21, Nikhef-2021-009",
    doi = "10.1140/epjc/s10052-021-09388-y",
    journal = "Eur. Phys. J. C",
    volume = "81",
    number = "7",
    pages = "638",
    year = "2021"
}

@article{Neubert:1993mb,
    author = "Neubert, Matthias",
    title = "{Heavy quark symmetry}",
    eprint = "hep-ph/9306320",
    archivePrefix = "arXiv",
    reportNumber = "SLAC-PUB-6263",
    doi = "10.1016/0370-1573(94)90091-4",
    journal = "Phys. Rept.",
    volume = "245",
    pages = "259--396",
    year = "1994"
}

@book{Manohar:2000dt,
    author = "Manohar, Aneesh V. and Wise, Mark B.",
    title = "{Heavy quark physics}",
    doi = "10.1017/9781009402125",
    isbn = "978-0-521-03757-0, 978-1-009-40212-5",
    volume = "10",
    year = "2000"
}

@article{Balitsky:1987bk,
    author = "Balitsky, I. I. and Braun, Vladimir M.",
    title = "{Evolution Equations for QCD String Operators}",
    reportNumber = "LENINGRAD-87-1351",
    doi = "10.1016/0550-3213(89)90168-5",
    journal = "Nucl. Phys. B",
    volume = "311",
    pages = "541--584",
    year = "1989"
}

@article{Isgur:1989vq,
    author = "Isgur, Nathan and Wise, Mark B.",
    title = "{Weak Decays of Heavy Mesons in the Static Quark Approximation}",
    reportNumber = "UTPT-89-27, CALT-68-1585",
    doi = "10.1016/0370-2693(89)90566-2",
    journal = "Phys. Lett. B",
    volume = "232",
    pages = "113--117",
    year = "1989"
}

@article{Eichten:1989zv,
    author = "Eichten, Estia and Hill, Brian Russell",
    title = "{An Effective Field Theory for the Calculation of Matrix Elements Involving Heavy Quarks}",
    reportNumber = "FERMILAB-PUB-89-184-T",
    doi = "10.1016/0370-2693(90)92049-O",
    journal = "Phys. Lett. B",
    volume = "234",
    pages = "511--516",
    year = "1990"
}

@article{Grinstein:1990mj,
    author = "Grinstein, Benjamin",
    title = "{The Static Quark Effective Theory}",
    reportNumber = "HUTP-90/A002",
    doi = "10.1016/0550-3213(90)90349-I",
    journal = "Nucl. Phys. B",
    volume = "339",
    pages = "253--268",
    year = "1990"
}

@article{Georgi:1990um,
    author = "Georgi, Howard",
    title = "{An Effective Field Theory for Heavy Quarks at Low-energies}",
    reportNumber = "HUTP-90/A007",
    doi = "10.1016/0370-2693(90)91128-X",
    journal = "Phys. Lett. B",
    volume = "240",
    pages = "447--450",
    year = "1990"
}

@article{Beneke:2005gs,
    author = "Beneke, M. and Yang, D.",
    title = "{Heavy-to-light B meson form-factors at large recoil energy: Spectator-scattering corrections}",
    eprint = "hep-ph/0508250",
    archivePrefix = "arXiv",
    reportNumber = "PITHA-05-10",
    doi = "10.1016/j.nuclphysb.2005.11.027",
    journal = "Nucl. Phys. B",
    volume = "736",
    pages = "34--81",
    year = "2006"
}

@article{Hill:2004if,
    author = "Hill, R. J. and Becher, T. and Lee, Seung J. and Neubert, M.",
    title = "{Sudakov resummation for subleading SCET currents and heavy-to-light form-factors}",
    eprint = "hep-ph/0404217",
    archivePrefix = "arXiv",
    reportNumber = "SLAC-PUB-10412, CLNS-04-1865",
    doi = "10.1088/1126-6708/2004/07/081",
    journal = "JHEP",
    volume = "07",
    pages = "081",
    year = "2004"
}

@article{Becher:2004kk,
    author = "Becher, Thomas and Hill, Richard J.",
    title = "{Loop corrections to heavy-to-light form-factors and evanescent operators in SCET}",
    eprint = "hep-ph/0408344",
    archivePrefix = "arXiv",
    reportNumber = "SLAC-PUB-10680",
    doi = "10.1088/1126-6708/2004/10/055",
    journal = "JHEP",
    volume = "10",
    pages = "055",
    year = "2004"
}

@article{Beneke:2004dp,
    author = "Beneke, M. and Feldmann, Th. and Seidel, D.",
    title = "{Exclusive radiative and electroweak $b \to  d$ and $b \to  s$ penguin decays at NLO}",
    eprint = "hep-ph/0412400",
    archivePrefix = "arXiv",
    reportNumber = "PITHA-04-19, SI-HEP-2004-13",
    doi = "10.1140/epjc/s2005-02181-5",
    journal = "Eur. Phys. J. C",
    volume = "41",
    pages = "173--188",
    year = "2005"
}

@article{Becher:2005fg,
    author = "Becher, Thomas and Hill, Richard J. and Neubert, Matthias",
    title = "{Factorization in B ---{\ensuremath{>}} V gamma decays}",
    eprint = "hep-ph/0503263",
    archivePrefix = "arXiv",
    reportNumber = "FERMILAB-PUB-05-032-T, SLAC-PUB-11045, CLNS-05-1913",
    doi = "10.1103/PhysRevD.72.094017",
    journal = "Phys. Rev. D",
    volume = "72",
    pages = "094017",
    year = "2005"
}

@article{Beneke:2000ry,
    author = "Beneke, M. and Buchalla, G. and Neubert, M. and Sachrajda, Christopher T.",
    title = "{QCD factorization for exclusive, nonleptonic B meson decays: General arguments and the case of heavy light final states}",
    eprint = "hep-ph/0006124",
    archivePrefix = "arXiv",
    reportNumber = "CERN-TH-2000-159, CLNS-00-1675, PITHA-00-06, SHEP-00-06",
    doi = "10.1016/S0550-3213(00)00559-9",
    journal = "Nucl. Phys. B",
    volume = "591",
    pages = "313--418",
    year = "2000"
}

@article{Beneke:2001ev,
    author = "Beneke, M. and Buchalla, G. and Neubert, M. and Sachrajda, Christopher T.",
    title = "{QCD factorization in B ---{\ensuremath{>}} pi K, pi pi decays and extraction of Wolfenstein parameters}",
    eprint = "hep-ph/0104110",
    archivePrefix = "arXiv",
    reportNumber = "CERN-TH-2001-107, CLNS-01-1728, PITHA-01-01, SHEP-01-11",
    doi = "10.1016/S0550-3213(01)00251-6",
    journal = "Nucl. Phys. B",
    volume = "606",
    pages = "245--321",
    year = "2001"
}

@article{Beneke:2003zv,
    author = "Beneke, Martin and Neubert, Matthias",
    title = "{QCD factorization for B ---{\ensuremath{>}} PP and B ---{\ensuremath{>}} PV decays}",
    eprint = "hep-ph/0308039",
    archivePrefix = "arXiv",
    reportNumber = "CLNS-03-1835, PITHA-03-06",
    doi = "10.1016/j.nuclphysb.2003.09.026",
    journal = "Nucl. Phys. B",
    volume = "675",
    pages = "333--415",
    year = "2003"
}

@article{Lu:2022kos,
    author = {L{\"u}, Cai-Dian and Shen, Yue-Long and Wang, Chao and Wang, Yu-Ming},
    title = "{Shedding new light on weak annihilation B-meson decays}",
    eprint = "2202.08073",
    archivePrefix = "arXiv",
    primaryClass = "hep-ph",
    doi = "10.1016/j.nuclphysb.2023.116175",
    journal = "Nucl. Phys. B",
    volume = "990",
    pages = "116175",
    year = "2023"
}

@article{DeFazio:2007hw,
    author = "De Fazio, Fulvia and Feldmann, Thorsten and Hurth, Tobias",
    title = "{SCET sum rules for B ---{\ensuremath{>}} P and B ---{\ensuremath{>}} V transition form factors}",
    eprint = "0711.3999",
    archivePrefix = "arXiv",
    primaryClass = "hep-ph",
    reportNumber = "CERN-PH-TH-2007-167, SLAC-PUB-12823, SI-HEP-2007-14, BARI-TH-07-582",
    doi = "10.1088/1126-6708/2008/02/031",
    journal = "JHEP",
    volume = "02",
    pages = "031",
    year = "2008"
}

@article{Wang:2015vgv,
    author = "Wang, Yu-Ming and Shen, Yue-Long",
    title = "{QCD corrections to B{\textrightarrow}{\ensuremath{\pi}} form factors from light-cone sum rules}",
    eprint = "1506.00667",
    archivePrefix = "arXiv",
    primaryClass = "hep-ph",
    reportNumber = "TUM-HEP-995-15",
    doi = "10.1016/j.nuclphysb.2015.07.016",
    journal = "Nucl. Phys. B",
    volume = "898",
    pages = "563--604",
    year = "2015"
}

@article{Wang:2017jow,
    author = {Wang, Yu-Ming and Wei, Yan-Bing and Shen, Yue-Long and L{\"u}, Cai-Dian},
    title = "{Perturbative corrections to B {\textrightarrow} D form factors in QCD}",
    eprint = "1701.06810",
    archivePrefix = "arXiv",
    primaryClass = "hep-ph",
    reportNumber = "UWTHPH-2016-29",
    doi = "10.1007/JHEP06(2017)062",
    journal = "JHEP",
    volume = "06",
    pages = "062",
    year = "2017"
}

@article{Lu:2018cfc,
    author = {L{\"u}, Cai-Dian and Shen, Yue-Long and Wang, Yu-Ming and Wei, Yan-Bing},
    title = "{QCD calculations of $B \to \pi, K$ form factors with higher-twist corrections}",
    eprint = "1810.00819",
    archivePrefix = "arXiv",
    primaryClass = "hep-ph",
    doi = "10.1007/JHEP01(2019)024",
    journal = "JHEP",
    volume = "01",
    pages = "024",
    year = "2019"
}

@article{Gao:2019lta,
    author = {Gao, Jing and L{\"u}, Cai-Dian and Shen, Yue-Long and Wang, Yu-Ming and Wei, Yan-Bing},
    title = "{Precision calculations of $B \to V$ form factors from soft-collinear effective theory sum rules on the light-cone}",
    eprint = "1907.11092",
    archivePrefix = "arXiv",
    primaryClass = "hep-ph",
    doi = "10.1103/PhysRevD.101.074035",
    journal = "Phys. Rev. D",
    volume = "101",
    number = "7",
    pages = "074035",
    year = "2020"
}

@article{Gao:2021sav,
    author = "Gao, Jing and Huber, Tobias and Ji, Yao and Wang, Chao and Wang, Yu-Ming and Wei, Yan-Bing",
    title = "{B {\textrightarrow} D{\ensuremath{\ell}}{\ensuremath{\nu}}$_{ℓ}$ form factors beyond leading power and extraction of |V$_{cb}$| and R(D)}",
    eprint = "2112.12674",
    archivePrefix = "arXiv",
    primaryClass = "hep-ph",
    reportNumber = "TUM-HEP-1331/21, SI-HEP-2021-37, P3H-21-103",
    doi = "10.1007/JHEP05(2022)024",
    journal = "JHEP",
    volume = "05",
    pages = "024",
    year = "2022"
}

@article{Cui:2022zwm,
    author = "Cui, Bo-Yan and Huang, Yong-Kang and Shen, Yue-Long and Wang, Chao and Wang, Yu-Ming",
    title = "{Precision calculations of B$_{d,s}$ {\textrightarrow} {\ensuremath{\pi}}, K decay form factors in soft-collinear effective theory}",
    eprint = "2212.11624",
    archivePrefix = "arXiv",
    primaryClass = "hep-ph",
    doi = "10.1007/JHEP03(2023)140",
    journal = "JHEP",
    volume = "03",
    pages = "140",
    year = "2023"
}

@article{Cui:2023jiw,
    author = "Cui, Bo-Yan and Huang, Yong-Kang and Wang, Yu-Ming and Zhao, Xue-Chen",
    title = "{Shedding new light on R(D(s)(*)) and |Vcb| from semileptonic B{\textasciimacron}(s){\textrightarrow}D(s)(*){\ensuremath{\ell}}{\ensuremath{\nu}}{\textasciimacron}{\ensuremath{\ell}} decays}",
    eprint = "2301.12391",
    archivePrefix = "arXiv",
    primaryClass = "hep-ph",
    doi = "10.1103/PhysRevD.108.L071504",
    journal = "Phys. Rev. D",
    volume = "108",
    number = "7",
    pages = "L071504",
    year = "2023"
}

@article{Khodjamirian:2023wol,
    author = "Khodjamirian, Alexander and Meli{\'c}, Bla{\v{z}}enka and Wang, Yu-Ming",
    title = "{A guide to the QCD light-cone sum rules for $b$-quark decays}",
    eprint = "2311.08700",
    archivePrefix = "arXiv",
    primaryClass = "hep-ph",
    reportNumber = "SI-HEP-2023-25, P3H-23-087; RBI-ThPhys-2023-37",
    doi = "10.1140/epjs/s11734-023-01046-6",
    journal = "Eur. Phys. J. ST",
    volume = "233",
    number = "2",
    pages = "271--298",
    year = "2024"
}

@article{Capdevila:2023yhq,
    author = "Capdevila, Bernat and Crivellin, Andreas and Matias, Joaquim",
    title = "{Review of semileptonic B anomalies}",
    eprint = "2309.01311",
    archivePrefix = "arXiv",
    primaryClass = "hep-ph",
    reportNumber = "PSI-PR-23-33, ZU-TH 50/23",
    doi = "10.1140/epjs/s11734-023-01012-2",
    journal = "Eur. Phys. J. ST",
    volume = "1",
    pages = "20",
    year = "2023"
}

@article{Qin:2022rlk,
    author = "Qin, Qin and Shen, Yue-Long and Wang, Chao and Wang, Yu-Ming",
    title = "{Deciphering the Long-Distance Penguin Contribution to B{\textasciimacron}d,s{\textrightarrow}{\ensuremath{\gamma}}{\ensuremath{\gamma}} Decays}",
    eprint = "2207.02691",
    archivePrefix = "arXiv",
    primaryClass = "hep-ph",
    doi = "10.1103/PhysRevLett.131.091902",
    journal = "Phys. Rev. Lett.",
    volume = "131",
    number = "9",
    pages = "091902",
    year = "2023"
}

@article{Braun:2012kp,
    author = "Braun, V. M. and Khodjamirian, A.",
    title = "{Soft contribution to $B\to \gamma \ell \nu_\ell$ and the $B$-meson distribution amplitude}",
    eprint = "1210.4453",
    archivePrefix = "arXiv",
    primaryClass = "hep-ph",
    reportNumber = "SI-HEP-2012-19",
    doi = "10.1016/j.physletb.2012.11.047",
    journal = "Phys. Lett. B",
    volume = "718",
    pages = "1014--1019",
    year = "2013"
}

@article{Wang:2016qii,
    author = "Wang, Yu-Ming",
    title = "{Factorization and dispersion relations for radiative leptonic $B$ decay}",
    eprint = "1606.03080",
    archivePrefix = "arXiv",
    primaryClass = "hep-ph",
    reportNumber = "UWTHPH-2016-10",
    doi = "10.1007/JHEP09(2016)159",
    journal = "JHEP",
    volume = "09",
    pages = "159",
    year = "2016"
}

@article{Wang:2018wfj,
    author = "Wang, Yu-Ming and Shen, Yue-Long",
    title = "{Subleading-power corrections to the radiative leptonic $B \to \gamma \ell \nu$ decay in QCD}",
    eprint = "1803.06667",
    archivePrefix = "arXiv",
    primaryClass = "hep-ph",
    doi = "10.1007/JHEP05(2018)184",
    journal = "JHEP",
    volume = "05",
    pages = "184",
    year = "2018"
}

@article{Galda:2022dhp,
    author = "Galda, Anne Mareike and Neubert, Matthias and Wang, Xing",
    title = "{Factorization and Sudakov resummation in leptonic radiative B decay {\textemdash} a reappraisal}",
    eprint = "2203.08202",
    archivePrefix = "arXiv",
    primaryClass = "hep-ph",
    reportNumber = "MITP/22-024",
    doi = "10.1007/JHEP07(2022)148",
    journal = "JHEP",
    volume = "07",
    pages = "148",
    year = "2022"
}

@article{Wang:2019msf,
    author = "Wang, Wei and Wang, Yu-Ming and Xu, Ji and Zhao, Shuai",
    title = "{$B$-meson light-cone distribution amplitude from Euclidean quantities}",
    eprint = "1908.09933",
    archivePrefix = "arXiv",
    primaryClass = "hep-ph",
    reportNumber = "JLAB-THY-19-3025",
    doi = "10.1103/PhysRevD.102.011502",
    journal = "Phys. Rev. D",
    volume = "102",
    number = "1",
    pages = "011502",
    year = "2020"
}

@article{Han:2024fkr,
    author = {Han, Xue-Ying and Hua, Jun and Ji, Xiangdong and L{\"u}, Cai-Dian and Wang, Wei and Xu, Ji and Zhang, Qi-An and Zhao, Shuai},
    title = "{Realistic method to access heavy meson light-cone distribution amplitudes from first-principle}",
    eprint = "2403.17492",
    archivePrefix = "arXiv",
    primaryClass = "hep-ph",
    doi = "10.1103/2t8s-w8t6",
    journal = "Phys. Rev. D",
    volume = "111",
    number = "11",
    pages = "L111503",
    year = "2025"
}

@article{LatticeParton:2024zko,
    author = "Han, Xue-Ying and others",
    collaboration = "Lattice Parton",
    title = "{Calculation of heavy meson light-cone distribution amplitudes from lattice QCD}",
    eprint = "2410.18654",
    archivePrefix = "arXiv",
    primaryClass = "hep-lat",
    doi = "10.1103/PhysRevD.111.034503",
    journal = "Phys. Rev. D",
    volume = "111",
    number = "3",
    pages = "034503",
    year = "2025"
}

@article{LPC:2026ffe,
    author = "Han, Xue-Ying and others",
    collaboration = "LPC",
    title = "{Continuum-Limit HQET LCDAs from Lattice QCD for Tightening B Decay Uncertainties}",
    eprint = "2605.10946",
    archivePrefix = "arXiv",
    primaryClass = "hep-lat",
    month = "4",
    year = "2026",
    doi = "",
    journal = "",
    volume = "",
    number = "",
    pages = "",
    year = ""
}

@article{LPC:2026vyv,
    author = "Gao, Hao-Fei and others",
    collaboration = "LPC",
    title = "{Determination of heavy meson light-cone distribution amplitudes: theoretical framework and lattice simulations}",
    eprint = "2604.25802",
    archivePrefix = "arXiv",
    primaryClass = "hep-lat",
    month = "4",
    year = "2026",
        doi = "",
    journal = "",
    volume = "",
    number = "",
    pages = "",
    year = ""
}

@article{Lee:2005gza,
    author = "Lee, Seung J. and Neubert, Matthias",
    title = "{Model-independent properties of the B-meson distribution amplitude}",
    eprint = "hep-ph/0509350",
    archivePrefix = "arXiv",
    reportNumber = "CLNS-05-1925",
    doi = "10.1103/PhysRevD.72.094028",
    journal = "Phys. Rev. D",
    volume = "72",
    pages = "094028",
    year = "2005"
}

@article{Feldmann:2014ika,
    author = "Feldmann, Thorsten and Lange, Bjorn O. and Wang, Yu-Ming",
    title = "{B -meson light-cone distribution amplitude: Perturbative constraints and asymptotic behavior in dual space}",
    eprint = "1404.1343",
    archivePrefix = "arXiv",
    primaryClass = "hep-ph",
    reportNumber = "SI-HEP-2013-18, QFET-2013-14, TTK-14-05, SFB-CPP-14-07, TUM-HEP-928-14",
    doi = "10.1103/PhysRevD.89.114001",
    journal = "Phys. Rev. D",
    volume = "89",
    number = "11",
    pages = "114001",
    year = "2014"
}

@article{Beneke:2023nmj,
    author = "Beneke, Martin and Finauri, Gael and Vos, K. Keri and Wei, Yanbing",
    title = "{QCD light-cone distribution amplitudes of heavy mesons from boosted HQET}",
    eprint = "2305.06401",
    archivePrefix = "arXiv",
    primaryClass = "hep-ph",
    reportNumber = "TUM-HEP-1455/23, Nikhef-2023-003",
    doi = "10.1007/JHEP09(2023)066",
    journal = "JHEP",
    volume = "09",
    pages = "066",
    year = "2023"
}

\clearpage
\onecolumngrid

\begingroup
\allowdisplaybreaks[2]
\renewcommand{\H}{H}
\providecommand{\qqquad}{\qquad\quad}
\providecommand{\qqqquad}{\qquad\qquad}
\providecommand{\dd}{}
\renewcommand{\dd}{ d}
\providecommand{\MS}{\overline{\mathrm{MS}}}
\providecommand{\cusp}{\Gamma_{\rm cusp}}

\setcounter{secnumdepth}{2}
\setcounter{section}{0}
\setcounter{equation}{0}
\setcounter{figure}{0}
\setcounter{table}{0}
\renewcommand{\thesection}{\arabic{section}}
\renewcommand{\thesubsection}{\thesection.\arabic{subsection}}
\renewcommand{\theequation}{S\arabic{equation}}
\renewcommand{\thefigure}{S\arabic{figure}}
\renewcommand{\thetable}{S\arabic{table}}
\renewcommand{\theHsection}{supplement.\arabic{section}}
\renewcommand{\theHsubsection}{\theHsection.\arabic{subsection}}
\renewcommand{\theHequation}{supplement.\arabic{equation}}
\renewcommand{\theHfigure}{supplement.\arabic{figure}}
\renewcommand{\theHtable}{supplement.\arabic{table}}

\makeatletter
\let\@hangfrom@section\@undefined
\renewcommand{\section}{\@startsection{section}{1}{\z@}%
  {-2.7ex \@plus -.6ex \@minus -.2ex}{1.2ex \@plus .2ex}%
  {\normalfont\large\bfseries\raggedright}}
\renewcommand{\subsection}{\@startsection{subsection}{2}{\z@}%
  {-2ex \@plus -.5ex \@minus -.2ex}{.8ex \@plus .2ex}%
  {\normalfont\normalsize\bfseries\raggedright}}
\def\p@subsection{}
\def\p@subsubsection{}
\makeatother

\newenvironment{supplementdisplay}
  {\par\addvspace{\medskipamount}\noindent\begin{minipage}{\linewidth}\centering}
  {\end{minipage}\par\addvspace{\medskipamount}}
\setlength{\tabcolsep}{6pt}
\setlength{\abovecaptionskip}{6pt}
\setlength{\belowcaptionskip}{4pt}

\begin{center}
  {\Large\bfseries Supplemental Material\par}
\end{center}

Here, we provide more detailed discussions that are supplementary to the main text. We first establish some conventions in Sec.~\ref{sec:sm-conventions} following the available literature closely. In Section~\ref{sec:sm-construction}
we elaborate the construction procedure of the Lange-Neubert kernel using conformal techniques in Sec.~\ref{sec:complete-kernel} with the full three-loop kernel presented in conformal, position, and momentum spaces. In Sec.~\ref{sec:sm-checks}, we describe consistency checks that validate our results and collect ancillary expressions.

\section{Definitions and conventions}
\label{sec:sm-conventions}
We use $a_s=\alpha_s/(4\pi)$ and $\mu\,dO/d\mu+\mathcal H O=0$ convention in the $\MS$ scheme. 
The color factors are $C_F=(N_c^2-1)/(2N_c)$, $C_A=N_c$, and $T_F=1/2$.
$n_f$ counts the number of active light flavors.
The twist-two light-heavy operator and the corresponding leading-twist $B$-meson LCDA are defined as~\cite{Grozin:1996pq,Lange:2003ff}
\begin{equation}
 O(z)=\bar q(zn)\,\slashed{n}\gamma_5[zn,0]h_v(0),
 \qquad n^2=0,\quad v^2=1,\quad n\cdot v=1\, .
 \label{eq:operator-definition}
\end{equation}
\begin{equation}
 \langle0|O(z)|\bar B(v)\rangle=iF_B(\mu)\Phi_+(z;\mu)\, ,
 \qquad
 \Phi_+(z;\mu)=\int_0^\infty\dd\omega\,\e^{-i\omega z}\phi_+(\omega;\mu)\, ,
 \qquad
 \phi_+(\omega;\mu) = \int_{-\infty}^{\infty}\frac{dz}{2\pi}\,\e^{i\omega z}\Phi_+(z;\mu)\, .
 \label{eq:lcda-definition}
\end{equation}
Here $[zn,0]$ is a straight lightlike Wilson line. The same state normalization defines
$\langle0|\bar q\gamma^\mu\gamma_5h_v|\bar B(v)\rangle=iF_Bv^\mu$.
Thus $F_B$ evolves independently of the light-heavy operator. The operator and decay constant obey
\begin{equation}
 \mu\frac{\dd O(z;\mu)}{\dd\mu}+[\mathcal H O](z;\mu)=0,
 \qquad \mu\frac{\dd F_B(\mu)}{\dd\mu}+\gamma_F F_B(\mu)=0.
 \label{eq:operator-current-evolution}
\end{equation}
Here $\gamma_F$ is the anomalous dimension of the HQET current defining $F_B$.
Dividing the matrix element by $F_B$ gives the normalized LCDA renormalization group equations (RGEs) in both position and momentum spaces,
\begin{align}
 &\mu\frac{\dd\Phi_+(z;\mu)}{\dd\mu}+[\mathcal H \Phi_+](z;\mu)-\gamma_F\Phi_+(z;\mu)=0\, ,
 \label{eq:lcda-evolution}
 \\
 &\mu\frac{\dd\phi_+(\omega;\mu)}{\dd\mu}
 +\int_0^\infty\dd\omega'\,
 \left[\widetilde{\mathcal H}(\omega,\omega';\mu)-\gamma_F(a_s)\delta(\omega-\omega')\right]\phi_+(\omega';\mu)=0\, .
 \label{eq:momentum-lcda-evolution}
\end{align}
The main task of this work is to determine the LN kernel in position (${\cal H}$) and momentum ($\widetilde{\cal H}(\omega,\omega';\mu)$) space.

\section{Construction}
\label{sec:sm-construction}
The two-loop conformal generator determines the nonlocal kernel.
Endpoint factorization fixes its additive constant.
\begin{center}
\begin{tikzpicture}[>=Stealth,box/.style={draw,rounded corners=2pt,align=center,inner sep=5pt,font=\small}]
\node[box,text width=34mm] (ll) at (0,0.85) {Known light-quark\\ quantum-corrected conformal generator};
\node[box,text width=28mm] (k) at (5.2,0.85) {$\mathcal K$ entering Eq.~\eqref{eq:conformal}};
\node[box,text width=34mm] (ad) at (0,-0.85) {known anomalous dimensions \\$\gamma_q,\gamma_h,\gamma_{\rm th}$};
\node[box,text width=28mm] (g) at (5.2,-0.85) {$\Gamma_+$ entering Eq.~\eqref{eq:conformal}};
\node[box,text width=29mm] (h) at (10.4,0) {Complete\\LN kernel};
\draw[->] (ll)--node[above,font=\small]{reduction}(k);
\draw[->] (k)--node[above,sloped,font=\small]{Mellin}node[below,sloped,font=\small]{$h^{(1)},h^{(2)}$}(h);
\draw[->] (ad)--node[above,font=\small]{matching}(g);
\draw[->] (g)--(h);
\end{tikzpicture}
\end{center}

\subsection{light-to-heavy reduction of the conformal generator}
\label{sec:sm-generator}
It is straightforward to apply the light-to-heavy reduction in the position space to obtain the quantum corrections to the special conformal generator in Eq.~\eqref{eq:conformal} from its flavor-nonsinglet twist-two light-light quark counterpart $\Delta_{qq}$ derived in~\cite{Braun:2016qlg}, which reads to two loops as,
\begin{align}
    \Delta_{qq}^{(1)} &= (z_1+z_2)\left(\beta_0+\frac{1}{2}\mathbb{H}^{(1)}\right)
   + z_{12}\Delta_{+}^{(1)}, \label{eq:Delta_qq1}\\
    \Delta_{qq}^{(2)}
    &= (z_1+z_2)\left(\beta_1+\frac{1}{2}\mathbb{H}^{(2)}\right)
   + \frac{1}{4}\left[\mathbb{H}^{(2)},\,z_1+z_2\right]
   + z_{12}\Delta_{+}^{(2)}\, ,\label{eq:Delta_qq2}
\end{align}
where ${\mathbb H}$ is the position-space Efremov--Radyushkin--Brodsky--Lepage (ERBL) kernel~\cite{Efremov:1979qk,Lepage:1979zb} fully known to three-loops~\cite{Balitsky:1987bk,Braun:2014vba,Chen:3LERBL}, and $\Delta_+$ is the so-called conformal anomaly currently known to the two-loop order, facilitating the translation of the position-space ERBL kernel from the so-called conformal scheme (CS), where $\Delta_+$ vanishes identically by definition, to the commonly used $\MS$-scheme. $\Delta_+$ can be systematically calculated through the diagrammatic procedure order by order in $a_s$. In the conformal approach, to obtain the $(\ell+1)$-loop ERBL kernel, the necessary ingredients are the $\ell$-loop conformal anomaly ($\Delta^{(\ell)}_+$) and ERBL kernel (${\mathbb H}^{(\ell)}$) and the $(\ell+1)$-loop DGLAP kernel. The ERBL kernel in position space takes the following form to all orders,
\begin{align}
[\mathbb{H} f](z_1,z_2)
=4\Bigg\{&
\mathbb{X} f(z_1,z_2)
+\int_0^1 d\alpha\,\frac{\bar{\alpha}}{\alpha}\,\tau(\alpha)
\bigl[2f(z_1,z_2)-f(z_{12}^{\alpha},z_2)-f(z_1,z_{21}^{\alpha})\bigr]\notag
\\
&+\int_0^1 d\alpha\int_0^{\bar{\alpha}}d\beta\,
\bigl[\chi(\alpha,\beta)+\chi^{\mathbb{P}}(\alpha,\beta)\mathbb{P}_{12}\bigr]
f(z_{12}^{\alpha},z_{21}^{\beta})
\Bigg\}\, ,
\end{align}
where $z_{12}^x = \bar x z_1+ x z_2$ with $\bar x = 1-x$, $\mathbb{X}$ is a number, and $f(z_1,z_2)$ serves as a test function. ${\mathbb P}_{12} f(x,y) = f(y,x)$ is the permutation operator. The conformal anomaly takes a similar form, 
\begin{align}
[\Delta_{+} f](z_1,z_2)
={}&
\int_0^1 d\alpha \int_0^{\bar{\alpha}} d\beta\,
\bigl[\omega(\alpha,\beta)
+\omega^{\mathbb{P}}(\alpha,\beta)\mathbb{P}_{12}\bigr]
\bigl[f(z_{12}^{\alpha},z_{21}^{\beta})
-f(z_{12}^{\beta},z_{21}^{\alpha})\bigr]
\notag\\
&+\int_0^1 du \int_0^1 dt\,\mathcal{X}(t)
\bigl[f(z_{12}^{ut},z_2)-f(z_1,z_{21}^{ut})\bigr]
\end{align}
to all orders in perturbation. So far, functions $\omega(\alpha,\beta),\omega^{\mathbb P}(\alpha,\beta),\mathcal{X}(t)$ have been diagrammatically calculated to two-loop as detailed in~\cite{Braun:2016qlg}. 

To obtain $\Delta$ in Eq.~\eqref{def:K}, first, we assemble all the known two-loop ingredients into Eqs.~\eqref{eq:Delta_qq1} and~\eqref{eq:Delta_qq2}. Next, we apply the light-to-heavy reduction procedure, which instructs us to eliminate all and only terms that relocate the heavy field, e.g., $f(z_{12}^\alpha,z_{21}^\beta)$ and $f(z_1,z_{12}^\alpha)$. Then the resulting expression is the quantum corrections to the special conformal generator $\Delta$~\eqref{def:K} of the light-heavy operator in Eq.~\eqref{eq:operator}. We note that individual contributions of ${\mathbb H}$ and $\Delta_+$ in Eqs.~\eqref{eq:Delta_qq1} and~\eqref{eq:Delta_qq2} can be divergent after performing the light-to-heavy reduction procedure, but the final expression of ${\cal K}$ is always well-defined. We confirm that the $O(a_s)$ correction to ${\cal K}$ obtained from the reduction procedure matches exactly with the diagrammatic result in~\cite{Braun:2019wyx}. 

Below, we present the explicit expression for the special conformal generator to two-loop order,
\begin{equation}
 \mathcal K=z\left[z\partial_z+2+a_s \Delta^{(1)}+a_s^2\Delta^{(2)}\right]+O(a_s^3)\, ,
 \label{eq:sm-generator}
\end{equation}
where 
\begin{align}
[\Delta^{(1)}f](z)&=(\beta_0+3C_F)f(z)+2C_F\int_0^1\dd u\left(\frac{2\bar u}{u}+\ln u\right)[f(z)-f(\bar u z)]\, ,
\label{eq:sm-generator-one-loop}
\\
[\Delta^{(2)}f](z)&=(\beta_1+2X_1+X_2)f(z)+\int_0^1\dd\alpha\,w_2(\alpha)[f(z)-f(\bar\alpha z)],
\quad w_2(\alpha)=(2-\alpha)\frac{\bar\alpha}{\alpha}\kappa(\alpha)-\widetilde\kappa(\alpha),
\label{eq:sm-generator-two-loop}
\end{align}
with 
\begin{align}
X_1 &=\frac{13}{12}C_F\beta_0
 +C_FC_A\left(6\zeta_3-\frac{2\pi^2}{3}+\frac{13}{6}\right)
 +2C_F^2\left(-6\zeta_3+\frac{\pi^2}{3}+\frac{21}{8}\right)\, ,\notag\\[0.5em]
X_2 &=\frac{47}{12}C_F\beta_0
 +C_FC_A\left(-12\zeta_3+\frac{\pi^2}{3}+\frac{17}{6}\right)
 +C_F^2\left(6\zeta_3+\frac{2\pi^2}{3}-\frac{53}{4}\right)\, ,\notag\\[0.5em]
\kappa(\alpha) &=C_F\beta_0\left(-\H_1+\frac53\right) - C_F C_A\frac13(\pi^2-4)   + 2 C_F^2\left(2\H_{1,1}+\frac{1+\bar\alpha}{\bar\alpha}\left(\H_{1,0}+\H_2\right) + \frac32\H_1\right)\, ,\notag\\[0.5em]
\widetilde\kappa(\alpha) &= 2C_F^2\bigg\{-\H_{1,0,0}-3\H_{2,1}-4\H_{2,0}-2\H_{1,2}-2\H_3+\left(1-\frac2\alpha+\alpha\right)\H_{1,1}+\left(6-\frac4\alpha-\alpha\right)\H_{1,0}\notag\\
&\qqqquad+(2+\alpha)\H_{0,0}-(5-3\alpha)\H_2+\left(3+\frac{\pi^2}{3}-\frac3\alpha\right)\H_1+\frac52\H_0+\frac{\pi^2}{6}(5-2\alpha)-\frac92\bar \alpha-6\zeta_3\bigg\}\notag\\
&+2 C_FC_A\bigg\{\H_{2,0}-\H_3+\left(3-\frac2\alpha-\alpha\right)(\H_2-\H_{1,0})+\left(\frac{\pi^2}3-\frac56\right)\H_0+\frac{\pi^2\bar \alpha-4}{3\alpha}-\alpha+3\zeta_3+\frac73\bigg\}\notag\\
&+2C_F\beta_0\bigg\{\H_2+\frac{\bar \alpha}{\alpha} \H_1-\frac{8}3\H_0-\frac{5}{3}\frac{\bar \alpha}\alpha-\frac{\pi^2}{6}\bigg\}\, ,
\end{align}
and the first two QCD $\beta$-function relevant for our study reads,
\begin{align}\label{eq:beta}
\beta_0 &= \frac{11}{3}C_A-\frac43T_F n_f\, ,\qqquad
\beta_1 =\frac23\left(17C_A^2-10 C_A  n_f T_F-6 C_F  n_f T_F \right)\, .
\end{align}

\section{Complete three-loop LN kernel in different representations}
\label{sec:complete-kernel}
The LN evolution kernel can be represented in conformal, position, and momentum spaces, with each representation advantageous in certain aspects. 

\subsection{Conformal-space representation}

The operator/logarithm part of the conformal representation to three loops can be easily recovered explicitly by collecting Eqs.~\eqref{eq:sm-generator}--\eqref{eq:beta} and substituting into Eq.~\eqref{eq:conformal}. 
Here the ``three-loop'' refers to the loop order of the constants $\cusp$ and $\Gamma_+$ with the ${\cal K}$ expanded to $O(a_s^2)$, consistent with conventional counting of loop calculations after transforming the conformal result to the position and momentum spaces. 
The constant part, i.e., $\Gamma_+$, explicitly reads,
\begin{align}
\Gamma_+ &= a_s (-5C_F) +a_s^2C_F\bigg\{C_F\left(-24\zeta_3+12\zeta_2-\frac32\right)+C_A\left(-6\zeta_3-\frac{43}{3}\zeta_2+\frac{67}{54}\right) +2n_f T_F\left[\frac{10}{3}\zeta_2+\frac{13}{27}\right]\bigg\} \notag\\
&+a_s^3C_F\bigg\{C_A^2\left(92\zeta_5+\frac{64}{3}\zeta_2\zeta_3+49\zeta_4-\frac{1162}9\zeta_3-\frac{14341}{81}\zeta_2+\frac{296647}{2916}\right) \notag\\
&\quad +C_A C_F\biggl(- 120 \zeta_5-16\zeta_2\zeta_3 +\frac{494}{3}\zeta_4-\frac{844}{3} \zeta_3 +\frac{410}{3}\zeta_2-\frac{151}{4}\biggr)\notag \\
&\quad+ C_F^2\left(240\zeta_5+32\zeta_2\zeta_3-144\zeta_4-68\zeta_3-18\zeta_2-\frac{29}2\right)\notag\\
&\quad+ C_A n_f T_F\left(-52\zeta_4+\frac{1016}{27}\zeta_3+\frac{9404}{81}\zeta_2-\frac{31390}{729}\right) + C_F n_f T_F\left(-\frac{184}{3}\zeta_4+\frac{832}9\zeta_3-\frac{28}3\zeta_2+\frac{521}{27}\right) \notag\\
&\quad +(n_f T_F)^2 \left(\frac{416}{27}\zeta_3-\frac{400}{27}\zeta_2+\frac{2212}{729}\right)\bigg\}+{\cal O}(a_s^4)\, ,\label{eq:Gamma_+}
\end{align}
which is obtained using the relation in Eq.~\eqref{eq:constant}. For completeness, we provide the anomalous dimensions  $\gamma_q$, $\gamma_h$, and $\gamma_{\rm th}$ in Eqs.~\eqref{eq:gq}--\eqref{eq:gs} to $O(a_s^3)$ necessary for determining $\Gamma_+$.

\subsection{Position-space representation}\label{sect:pos-LN-kernel}

The position-space LN kernel takes the all-order form as in Eq.~\eqref{eq:kernel}. To three loops, we have,
\begin{align}
[\mathcal H f](z)={}&\cusp(a_s)\left\{\ln(i\mu e^{\gamma_E}z)f(z)
+\int_0^1\dd\alpha\,\frac{\bar\alpha}{\alpha}
[1+a_s h^{(1)}(\alpha)+a_s^2h^{(2)}(\alpha)][f(z)-f(\bar\alpha z)]\right\}
+\gamma_+(a_s)f(z) + O(a_s^4)\, ,\label{eq:complete-position-kernel}
\end{align}
where
\begin{align}
\gamma_+(a_s)={}&-a_s C_F+a_s^2 C_F\left\{4C_F\left(\frac{21}{8}+\frac{\pi^2}{3}-6\zeta_3\right)
+C_A\left(\frac{83}{9}-\frac{2\pi^2}{3}-6\zeta_3\right)+\beta_0\left(\frac{35}{18}-\frac{\pi^2}{6}\right)\right\}\notag\\
&+a_s^3C_F\bigg\{C_A^2\left(92\zeta_5+\frac{64}3\zeta_2\zeta_3+\frac{191}3\zeta_4-\frac{1382}9\zeta_3-\frac{9097}{81}\zeta_2+\frac{479167}{2916}\right)\notag\\
&\quad+C_FC_A\left(-240\zeta_5+48\zeta_2\zeta_3+176\zeta_4-\frac{1804}9\zeta_3+\frac{1024}9\zeta_2-\frac{2039}{36}\right)\notag\\
&\quad+C_F^2\left(320\zeta_5-64\zeta_2\zeta_3-114\zeta_4-72\zeta_3-28\zeta_2+\frac{127}2\right)\notag\\
&\quad+2C_AT_Fn_f\left(-\frac{38}3\zeta_4+\frac{28}{27}\zeta_3+\frac{1702}{81}\zeta_2-\frac{16451}{729}\right)+2C_FT_Fn_f\left(-32\zeta_4+64\zeta_3-\frac{94}9\zeta_2-\frac{907}{54}\right)\notag\\
&\quad+4(T_Fn_f)^2\left(\frac{56}{27}\zeta_3-\frac{20}{27}\zeta_2-\frac{743}{729}\right)\bigg\}\, ,
\end{align}
and
\begin{align}
 h^{(1)}(\alpha)&=\ln\bar\alpha\left[\beta_0+2C_F\left(\ln\bar\alpha-\frac{1+\bar\alpha}{\bar\alpha}\ln\alpha-\frac32\right)\right]\, ,\notag\\
 h^{(2)}(\alpha)&=\beta_1\ln\bar\alpha+\frac12\beta_0^2\ln^2\bar\alpha
       +C_F\beta_0 F_\beta(\alpha)+C_FC_AF_A(\alpha)+C_F^2F_F(\alpha)\, ,
 \label{eq:h1-h2}
\end{align}
with
\begin{align}
F_\beta(\alpha)={}&\ln\bar\alpha\left[\frac32\ln^2\bar\alpha-\left(2\ln\alpha+\frac43\right)\ln\bar\alpha-\frac12
 +\frac{2\operatorname{Li}_2(\alpha)}{\bar\alpha}
 -\frac{2-\alpha}{3\bar\alpha}(10\ln\alpha+\pi^2)\right],
 \label{eq:weight-Fbeta}\\
 F_A(\alpha)={}&\ln\bar\alpha\left[-1+\frac{1-\pi^2}{3}\ln\bar\alpha
 +\frac{2(2-\alpha)}{\bar\alpha}  \left(\ln\alpha\operatorname{Li}_2(\alpha)-3\operatorname{Li}_3(\alpha)+3\zeta_3\right)
 +\frac{2\pi^2(2-\alpha)+11\alpha-16}{3\bar\alpha}\ln\alpha\right]\, ,
 \notag\\
 F_F(\alpha)={}&\frac{4(2-\alpha)}{\bar\alpha}\Big[2H_{1,1,0,0}(\alpha)+H_{1,1,2}(\alpha)+H_{1,2,0}(\alpha)
 +H_{1,2,1}(\alpha)+H_{1,3}(\alpha)\Big]
 +\ln^4\bar\alpha+\left(\frac{3\alpha-2}{\bar\alpha}\ln\alpha-\frac92\right)\ln^3\bar\alpha\notag\\
 &+\left[\frac{25}{2}+6\ln\alpha
 +\frac{2-\alpha}{\bar\alpha}
   \left(\ln^2\alpha-\pi^2+2\operatorname{Li}_2(\bar\alpha)\right)
 +\frac{4\operatorname{Li}_2(\alpha)}{\bar\alpha}\right]\ln^2\bar\alpha\notag\\
 &+\left[-\frac32-\frac{\alpha}{\bar\alpha}\ln\alpha
 +\frac{2-\alpha}{\bar\alpha}
  \left(\pi^2+10\operatorname{Li}_3(\alpha)-2\operatorname{Li}_3(\bar\alpha)
       -10\zeta_3-6\ln\alpha\operatorname{Li}_2(\alpha)\right)
 -\frac{6\operatorname{Li}_2(\alpha)}{\bar\alpha}\right]\ln\bar\alpha.
    \label{eq:h2-parts}
\end{align}

\subsection{Momentum-space representation}
\label{sec:momentum-kernel}
Replacing the position space test function $f(z)$ with its momentum space counterpart following the definition in Eq.~\eqref{eq:lcda-definition}, and then multiplying both sides by $\e^{i\omega z}$ and integrating over $z$ allows us to obtain the momentum space LN kernel $\widetilde{\cal H}(\omega,\omega';\mu)$, which enters the momentum space RGE of the leading-twist $B$-meson LCDA $\phi_+(\omega;\mu)$ in the form of Eq.~\eqref{eq:momentum-lcda-evolution}. Writing the momentum-space LN kernel as,
\begin{align}
 \widetilde{\mathcal H}(\omega,\omega';\mu)={}&
 \left[\cusp(a_s)\ln\frac{\mu}{\omega}+\Gamma_+(a_s)\right]\delta(\omega-\omega')\\
 &-\cusp(a_s)\left[ L_+(\omega,\omega')
       +a_s\widetilde h^{(1)}(\omega,\omega')+a_s^2\widetilde h^{(2)}(\omega,\omega')\right]
 +O(a_s^4)\, ,
\end{align}
where
\begin{align}
    L_+(\omega,\omega')=\left(\frac{\theta(\omega-\omega')}{\omega-\omega'}
 +\frac{\omega\,\theta(\omega'-\omega)}{\omega'(\omega'-\omega)}\right)_{+}\, ,
\end{align}
with the plus-prescription for an arbitrary test function $f(\omega)$ under the $\displaystyle \int^\infty_0 d\omega'$ integral in Eq.~\eqref{eq:momentum-lcda-evolution} defined as,
\begin{equation}
 \int_0^\infty\dd\omega'\, L_+(\omega,\omega')f(\omega')
 \equiv\int_0^\infty\dd\omega'\,\left(\frac{\theta(\omega-\omega')}{\omega-\omega'}
 +\frac{\omega\,\theta(\omega'-\omega)}{\omega'(\omega'-\omega)}\right)
 [f(\omega')-f(\omega)] \, .
 \label{eq:momentum-plus-action}
\end{equation}
%
The nonlocal corrections $\widetilde h^{(1)}(\omega,\omega')\, , \widetilde h^{(2)}(\omega,\omega')$ associated with $h(\alpha,a_s)$ in Eq.~\eqref{eq:kernel} read
\begin{equation}
 \widetilde h^{(\ell)}(\omega,\omega')=
 \frac{\omega\,\theta(\omega'-\omega)}{\omega'(\omega'-\omega)}
 h^{(\ell)}\!\left(1-\frac{\omega}{\omega'}\right)\, ,
 \label{eq:momentum-regular-kernels}
\end{equation}
where $h^{(1)}$ and $h^{(2)}$ are given in Eq.~\eqref{eq:h1-h2}. The simple relation above between the momentum space $\widetilde h^{(\ell)}$ and position space $h^{(\ell)}$ holds to all orders.

\subsection{Anomalous dimensions}\label{sect:ads}
We follow the following convention for the $a_s$-expansion of the cusp anomalous dimension,
\begin{align}
 \cusp(a_s)&=\sum_{n=0}^{2}a_s^{n+1}\Gamma_{\rm cusp}^{(n)}\, ,\label{eq:as-expansion}
 \end{align}
 and to obtain the LN kernel to the $a_s^3$ order, we need~\cite{Korchemsky:1987wg,Moch:2004pa},
 \begin{align}
 \cusp^{(0)}&=4C_F,\\
 \cusp^{(1)}&=4C_F\left[C_A\left(\frac{67}{9}-\frac{\pi^2}{3}\right)-\frac{20}{9}T_Fn_f\right],\\
 \cusp^{(2)}&=4C_F\bigg[C_A^2\left(\frac{245}{6}-\frac{134\pi^2}{27}+\frac{11\pi^4}{45}+\frac{22}{3}\zeta_3\right)
 +C_AT_Fn_f\left(-\frac{418}{27}+\frac{40\pi^2}{27}-\frac{56}{3}\zeta_3\right)\notag\\
 &\hspace{17mm}+C_FT_Fn_f\left(-\frac{55}{3}+16\zeta_3\right)-\frac{16}{27}(T_Fn_f)^2\bigg]\, .
 \label{eq:cusp-coefficients}
\end{align}
The anomalous dimension of the HQET decay constant $F_B(\mu)$ is required for the scale evolution of the twist-two $B$-meson LCDA $\phi_+$, hence the RG-consistency check via the three-loop bare jet function. Following the same notational convention as in Eq.~\eqref{eq:as-expansion}, we quote~\cite{Chetyrkin:2003vi}, 
\begin{align}
 \gamma_{F}^{(0)}={}&-3C_F,\notag\\
 \gamma_{F}^{(1)}={}&C_F\left[C_F\left(-16\zeta_2+\frac52\right)+C_A\left(4\zeta_2-\frac{49}{6}\right)+\frac{10}{3}T_Fn_f\right],\notag\\
 \gamma_{F}^{(2)}={}&C_F\bigg[C_F^2\left(-80\zeta_4-36\zeta_3+64\zeta_2-\frac{37}{2}\right)
 +C_FC_A\left(-16\zeta_4+\frac{142}{3}\zeta_3-\frac{1184}{9}\zeta_2-\frac{655}{36}\right)\notag\\
 &\quad+C_A^2\left(-24\zeta_4-\frac{22}{3}\zeta_3+\frac{260}{9}\zeta_2+\frac{1451}{108}\right)
 +C_FT_Fn_f\left(-\frac{176}{3}\zeta_3+\frac{448}{9}\zeta_2+\frac{470}{9}\right)\notag\\
 &\quad+C_AT_Fn_f\left(\frac{152}{3}\zeta_3-\frac{112}{9}\zeta_2-\frac{512}{27}\right)+\frac{140}{27}(T_Fn_f)^2\bigg].
 \label{eq:gammaF-coefficients}
\end{align}
The light-quark, heavy-quark, and threshold anomalous dimensions to the $a_s^3$ order read~\cite{Moch:2005id,Grozin:2014hna,Li:2014afw}
\begin{align}
\gamma_q &= a_s(-3C_F) + a_s^2C_F\left[C_A\left(26\zeta_3-\frac{961}{54}-11\zeta_2\right)+C_F\left(-24\zeta_3+12\zeta_2-\frac32\right)+n_fT_F\left(4\zeta_2+\frac{130}{27}\right)\right]\notag\\
&+a_s^3C_F\bigg[C_A^2 \biggl(\frac{3526 \zeta_3}{9}-\frac{44 \pi ^2 \zeta_3}{9}
-136 \zeta_5 -\frac{139345}{2916}-\frac{7163 \pi ^2}{486}
-\frac{83 \pi^4}{90}\biggr)\notag\\
&\quad +C_A C_F\biggl(-\frac{844 \zeta_3}{3}
-\frac{8 \pi ^2 \zeta_3}{3} - 120 \zeta_5 -\frac{151}{4}
+\frac{205 \pi^2}{9}+\frac{247 \pi ^4}{135}\biggr)\notag \\
&\quad +C_F^2 \biggl(-68 \zeta_3+\frac{16 \pi ^2 \zeta_3}{3}+240 \zeta_5
-\frac{29}{2}-3 \pi^2-\frac{8 \pi^4}{5}\biggr) +C_A n_f T_F \biggl(-\frac{1928 \zeta_3}{27}-\frac{17318}{729}
+\frac{2594 \pi ^2}{243}+\frac{22 \pi^4}{45}\biggr) \notag\\
&\quad +C_F n_f T_F \biggl(\frac{512 \zeta_3}{9}+\frac{2953}{27}
-\frac{26\pi^2}{9}-\frac{28 \pi ^4}{27}\biggr)  +(n_f T_F)^2 \biggl(-\frac{32 \zeta_3}{27}+\frac{9668}{729}
-\frac{40 \pi^2}{27}\biggr)\bigg]\, ,\label{eq:gq}
\\[0.8em]
\gamma_h &= a_s(-2C_F)+a_s^2C_F\left[C_A\left(-4\zeta_3+4\zeta_2-\frac{98}{9}\right)+n_f T_F\frac{40}{9}\right] \notag\\
&+ a_s^3C_F \bigg[ C_A^2\biggl(-\frac{4}{3} \pi ^2 \zeta _3-\frac{740 \zeta_3}{9}+36 
\zeta _5-\frac{22 \pi ^4}{45}+\frac{304 \pi ^2}{27}-\frac{343}{9}\biggr) 
+C_F n_f T_F \biggl(\frac{110}{3}-32\zeta _3\biggr)  \notag\\
&\quad+ C_A n_f T_F \biggl(\frac{496 \zeta _3}{9}-\frac{80 \pi 
^2}{27}+\frac{356}{27}\biggr)
+ \frac{32}{27}(n_f T_F)^2 \bigg]\, ,\label{eq:gh} 
\\[0.8em]
\gamma_{\rm th} &= a_s(0) + a_s^2 C_F\left[C_A\left(28\zeta_3+\frac{22}{3}\zeta_2-\frac{808}{27}\right)+n_fT_F\left(-\frac{8}{3}\zeta_2+\frac{224}{27}\right)\right]\notag\\
& + a_s^3C_F\bigg[C_A^2\left(-\frac{176}{3}\zeta_2\zeta_3+\frac{12650}{81}\zeta_2+\frac{1316}3\zeta_3-176\zeta_4-192\zeta_5-\frac{136781}{729}\right)\notag\\
&\quad+C_An_fT_F\left(-\frac{5656}{81}\zeta_2-\frac{1456}{27}\zeta_3+96\zeta_4+\frac{23684}{729}\right)+C_Fn_fT_F\left(-8\zeta_2-\frac{608}9\zeta_3-32\zeta_4+\frac{3422}{27}\right)\notag\\
&\quad+\left(n_fT_F\right)^2\left(\frac{160}{27}\zeta_2-\frac{448}{27}\zeta_3+\frac{8320}{729}\right)\bigg]\, .\label{eq:gs}
\end{align}
Using Eq.~\eqref{eq:constant}, they yield the explicit expression of $\Gamma_+$ in Eq.~\eqref{eq:Gamma_+}. 
\subsection{ERBL consistency}
After obtaining the LN kernel, we compare it with the light-to-heavy limit
of its companion three-loop ERBL kernel, whose complete form in the position space has recently been derived by some of the authors of this Letter~\cite{Chen:3LERBL}. 
Denote $\mathbb H^{(2)}$ as the three-loop ERBL kernel, the light-to-heavy reduction that maps ${\mathbb H}^{(2)}$ to the three-loop LN kernel ${\cal H}^{(2)}$ can be performed by acting ${\mathbb H}^{(2)}$ onto the test function $g(z_1,z_2) \equiv f(z_1)\e^{-iMz_2}$,
\begin{equation}
 \begin{aligned}
 \lim_{M\to\infty}\Bigg\{
 &\left[\mathbb H^{(2)}\!\left(f(z_1)\e^{-iMz_2}\right)\right]_{z_1=z,\,z_2=0}
 -\Gamma^{(2)}_{\rm cusp}\ln\frac M\mu\,f(z)\Bigg\}
 =[\mathcal H^{(2)}f](z)+(\gamma_{q}^{(2)}-\gamma_{h}^{(2)})f(z)\, ,
 \end{aligned}
 \label{eq:erbl-coordinate-comparison}
\end{equation}
where $M$ is the auxiliary heavy quark mass. The replacement rule $z_1=z$ is understood as $z_1\to {\rm Re}(z)-i0\equiv z$,  i.e., $z$ is a complex number in the lower half of the complex plane by the convergence requirement in the light-to-heavy reduction and is also consistent with the analyticity condition of the $B$-meson LCDAs~\cite{Braun:2017liq}.  The subtracted logarithm is the matching coefficient between QCD heavy quark and HQET heavy-quark field $h_v$. The remaining local difference uses the anomalous dimensions of the light- and heavy-quark form factors consistent with discussions in section ``The local constant from Wilson-line geometry'' of the main text, and in Ref.~\cite{Beneke:2024cpq}.

\section{Bare jet renormalization constant and anomalous dimension}
\label{sec:sm-checks}

Scale invariance of the leading-power $B$-meson decay form factors $F^{\rm (LP)}_{V,A}$ together with the factorization formula in Eq.~\eqref{eq:lp-factorization}, with $f_B$ begin the scale-independent QCD
decay constant, dictates the following relation among the renormalization constants of various quantities in Eq.~\eqref{eq:lp-factorization},
\begin{equation}
 Z_H Z_F^{-1}Z_O^T Z_J=\mathbf1,\qquad
 Z_J=Z_H^{-1}Z_F(Z_O^T)^{-1}\, ,
 \label{eq:jet-counterterm-factorization}
\end{equation}
where $Z_H$, $Z_F$, and $Z_O$ are the renormalization constants of the hard function, the $B$-meson decay constant in HQET, and the twist-two light-heavy operator, respectively. They are defined as,
\begin{equation}
 \begin{aligned}
 H&=Z_H H_{\rm bare},\qquad F_B=Z_F F_{B,\rm bare}\, ,\qquad
 \widetilde O=Z_O\otimes \widetilde O_{\rm bare},\qquad
 J_\gamma=Z_J\otimes J_\gamma^{\rm bare}.
 \end{aligned}
 \label{eq:jet-factor-counterterms}
\end{equation}
Here $Z_H$ and $Z_F$ are simple constants, while $Z_O$ and $Z_J$ are convolution
operators. $\widetilde O\equiv\widetilde O(\omega)$ is the momentum space light-heavy operator defined by Fourier transforming Eq.~\eqref{eq:operator} via Eq.~\eqref{eq:lcda-definition}.
The transpose convolution in Eq.~\eqref{eq:jet-counterterm-factorization} is defined from $Z_O$ as follows,
\begin{equation}
 [Z_O^{T} f](\omega')=\omega'\int_0^\infty\frac{\dd\omega}{\omega}
 \,f(\omega)Z_O(\omega,\omega',\mu)\, ,
 \label{eq:sm-transpose-action}
\end{equation}
with the evolution kernels defined in the same fashion. 
%
In terms of anomalous dimensions with $\dd H/\dd\ln\mu=\gamma_HH$ and
$\dd J/\dd\ln\mu=-\gamma_J\otimes J$, scale independence of $F_{V,A}^{(\rm LP)}$ implies
\begin{equation}
 \begin{aligned}
 \gamma_J=(\gamma_H+\gamma_F)\mathbf1-\mathcal H^T, \qquad 
 \gamma_H=\cusp\ln(2E_\gamma/\mu)+\gamma_q+\gamma_h-\gamma_F\, ,
 \end{aligned}
 \label{eq:sm-jet-consistency}
\end{equation}
and putting all ingredients together, we have,
\begin{equation}
 \gamma_J=\cusp(a_s)\left[
\ln(2E_\gamma\omega/\mu^2)\mathbf1+L_+^T+a_s\widetilde h^{(1)T}
 +a_s^2\widetilde h^{(2)T}\right]
 +\gamma_{\rm th}(a_s)\mathbf1+O(a_s^4).
 \label{eq:jet-anomalous-coefficients}
\end{equation}
The transposed kernels acting on an auxiliary function $f(\omega)$ takes the form,
\begin{equation}
 \bigl[L_+^T f\bigr](\omega)
 =\int_0^1\frac{\dd\alpha}{\alpha}
 \left[f(\bar\alpha\omega)+f\!\left(\frac{\omega}{\bar\alpha}\right)
       -2f(\omega)\right]\, , \qquad
 \bigl[\widetilde h^{(\ell)T}f\bigr](\omega)
 =\int_0^1\frac{\dd\alpha}{\alpha}
 \,h^{(\ell)}(\alpha)f(\bar\alpha\omega)\, .
 \label{eq:jet-regular-action}
\end{equation}
The functions $h^{(\ell)}(\alpha)$ and the cusp and threshold anomalous
dimensions are given in Secs.~\ref{sect:pos-LN-kernel} and~\ref{sect:ads}, respectively.

Expanding the evolution kernel/anomalous dimension of the jet function and its renormalization constant in the standard form,
\begin{equation}
 \gamma_J=\sum_{n=0}^\infty a_s^{n+1}\gamma^{(n)}_J,\qquad
 Z_J=\sum_{n=0}^\infty a_s^nZ_{J}^{(n)} \, , \quad \text{with} \quad Z_J^{(0)} = \mathbf1
 \label{eq:jet-z-definition}
\end{equation}
By scale-independence of the bare quantities, the jet evolution kernel/anomalous dimension can be easily derived from its renormalization constant, 
$\dd Z_J/\dd\ln\mu=-\gamma_J\otimes Z_J$~\cite{Liu:2020ydl}.
using 
$\dd a_s/\dd\ln\mu=-2\epsilon a_s-2\beta_0a_s^2-2\beta_1a_s^3+\cdots$ in $d=4-2\epsilon$ dimensions.
Due to the double-pole structure in $Z_J$, which is described by the cusp anomalous dimension, $\gamma_J^{(n)}$ is scale-dependent and satisfies,
\begin{equation*}
 \frac{\partial}{\partial\ln\mu}\gamma_J^{(n)}
 =-2\Gamma_{\rm cusp}^{(n)}\mathbf1\, .
\end{equation*}
Comparing powers of $a_s$ gives us recurrence relation,
\begin{equation}
 \left(2n\epsilon-\frac{\partial}{\partial{\ln\mu}}\right)Z_{J}^{(n)}
 =\sum_{m=0}^{n-1}\gamma_J^{(m)} Z_{J}^{(n-1-m)}
 -2\sum_{k=0}^{n-2}(n-k-1)\beta_k Z_{J}^{(n-k-1)}\, ,
 \label{eq:jet-z-recurrence}
\end{equation}
where the second sum vanishes for $n=1$. Using the renormalizability of the jet function, the coefficients of the higher-order poles $1/\epsilon^6\, ,\cdots 1/\epsilon^3$ are completely determined by the lower $a_s$-order anomalous dimensions of the jet function, the cusp anomalous dimensions, and the QCD $\beta$-function. Below, we give the explicit expression up to $Z_J^{(3)}$, 
\begin{align}
 Z_{J}^{(1)}={}&-\frac{\Gamma_{\rm cusp}^{(0)}}{2\epsilon^2}+\frac{\gamma_J^{(0)}}{2\epsilon},
 \label{eq:jet-z-one-loop}\\
 Z_{J}^{(2)}={}&\frac{(\Gamma_{\rm cusp}^{(0)})^2}{8\epsilon^4}
 -\frac{\Gamma_{\rm cusp}^{(0)}(2\gamma_J^{(0)}-3\beta_0)}{8\epsilon^3}
 +\frac{(\gamma_J^{(0)})^2-2\beta_0\gamma_J^{(0)}-\Gamma_{\rm cusp}^{(1)}}{8\epsilon^2}
 +\frac{\gamma_J^{(1)}}{4\epsilon},
 \label{eq:jet-z-two-loop}\\
 Z_{J}^{(3)}={}&-\frac{(\Gamma_{\rm cusp}^{(0)})^3}{48\epsilon^6}
 +\frac{(\Gamma_{\rm cusp}^{(0)})^2(\gamma_J^{(0)}-3\beta_0)}{16\epsilon^5}
 +\frac{\Gamma_{\rm cusp}^{(0)}(-9(\gamma_J^{(0)})^2+45\beta_0\gamma_J^{(0)}+9\Gamma_{\rm cusp}^{(1)}-44\beta_0^2)}{144\epsilon^4}\notag\\
 &+\frac{(\gamma_J^{(0)})^3-6\beta_0(\gamma_J^{(0)})^2+(8\beta_0^2-3\Gamma_{\rm cusp}^{(1)})\gamma_J^{(0)}-6\Gamma_{\rm cusp}^{(0)}\gamma_J^{(1)}}{48\epsilon^3}
 +\frac{5\beta_0\Gamma_{\rm cusp}^{(1)}+8\beta_1\Gamma_{\rm cusp}^{(0)}}{36\epsilon^3}\notag\\
 &+\frac{(\gamma_J^{(0)})\gamma_J^{(1)}+2\gamma_J^{(1)}(\gamma_J^{(0)})}{24\epsilon^2}
 -\frac{3\beta_0\gamma_J^{(1)}+3\beta_1(\gamma_J^{(0)})+\Gamma_{\rm cusp}^{(2)}}{18\epsilon^2}
 +\frac{\gamma_J^{(2)}}{6\epsilon}\, ,
 \label{eq:jet-z-three-loop}
\end{align}
where we note that since $\gamma_J^{(n)}$ are integral operators, $\gamma_J^{(n)}$ and $\gamma_J^{(m)}$ do not commute for $n\neq m$, e.g., 
\begin{equation}
 \Big[\big(\gamma_J^{(1)}\gamma_J^{(0)}-\gamma_J^{(0)}\gamma_J^{(1)}\big)f\Big](\omega)
 =(\Gamma_{\rm cusp}^{(0)})^2\int_0^1\frac{\dd\alpha}{\alpha}
 \,h^{(1)}(\alpha)\ln\bar\alpha\,f(\bar\alpha\omega)\, ,
 \label{eq:jet-z-commutator}
\end{equation}
which is a direct consequence of the conformal representation of the LN kernel in Eq.~\eqref{eq:conformal}. 
The cancellation of all $\epsilon$-pole in the independently calculated three-loop jet
function provides highly nontrivial validation to our three-loop LN kernel presented in sect.~\ref{sec:complete-kernel}. See our companion paper~\cite{Huang:3Ljet} for explicit expression of $\gamma_J$ and $Z_J$.


\label{sm:core-end}

\ifdefined\LNIncludeResearchNotes
 \clearpage
 \setcounter{section}{0}
 \setcounter{equation}{0}
 \setcounter{figure}{0}
 \setcounter{table}{0}
 \renewcommand{\thesection}{R\arabic{section}}
 \renewcommand{\theequation}{R\arabic{equation}}
 \renewcommand{\thefigure}{R\arabic{figure}}
 \renewcommand{\thetable}{R\arabic{table}}
 \renewcommand{\theHsection}{research.\arabic{section}}
 \renewcommand{\theHsubsection}{\theHsection.\arabic{subsection}}
 \renewcommand{\theHequation}{research.\arabic{equation}}
 \renewcommand{\theHfigure}{research.\arabic{figure}}
 \renewcommand{\theHtable}{research.\arabic{table}}
 \begin{center}
  {\Large\bfseries Additional Research Notes\par}
  \smallskip
  For collaborator review; not part of the Supplemental Material
 \end{center}
 \phantomsection
 \label{research:begin}
 \input{research-notes.tex}
\fi
\endgroup

\end{document}